# Parcels and particles: Markov blankets in the brain

*Karl J. Friston[1], Erik D. Fagerholm[2], Tahereh S. Zarghami[3], Thomas Parr[1], Inês Hipólito[4], Loïc Magrou[5] and Adeel Razi[1,6\*]*

[1]*The Wellcome Centre for Human Neuroimaging, University College London, Queen Square, London WC1N 3AR. UK*

[2]*Department of Neuroimaging, King's College London, SE5 8AF, UK*

[3]*Bio-Electric Department, School of Electrical and Computer Engineering, University of Tehran, Amirabad, Tehran, Iran*

[4] *Faculty of Arts, Social Sciences and Humanities, University of Wollongong, Wollongong, Australia*

[5] *INSERM U1208, Stem-cell and Brain Research Institute, 69675 Bron Cedex, France*

[6]*Turner Institute for Brain and Mental Health, Monash University, Clayton, Australia*

\* Correspondence to: Adeel Razi (adeel.razi@monash.edu)

**E-mails**: Karl Friston k.friston@ucl.ac.uk; Erik Fagerholm erik.fagerholm@kcl.ac.uk; Tahereh Zarghami tzarghami@ut.ac.ir; Thomas Parr thomas.parr.12@ucl.ac.uk; Ines Hipolito inesh@uow.edu.au; Loïc Magrou loic.magrou@gmail.com; Adeel Razi adeel.razi@monash.edu

## Abstract

At the inception of human brain mapping, two principles of functional anatomy underwrote most conceptions – and analyses – of distributed brain responses: namely functional *segregation* and *integration*. There are currently two main approaches to characterising functional integration. The first is a mechanistic modelling of connectomics in terms of directed *effective* connectivity that mediates neuronal message passing and dynamics on neuronal circuits. The second phenomenological approach usually characterises undirected *functional connectivity* (i.e., measurable correlations), in terms of intrinsic brain networks, self-organised criticality, dynamical instability, *etc*. This paper describes a treatment of effective connectivity that speaks to the emergence of intrinsic brain networks and critical dynamics. It is predicated on the notion of *Markov blankets* that play a fundamental role in the self-organisation of far from equilibrium systems.

Using the apparatus of the *renormalisation group*, we show that much of the phenomenology found in network neuroscience is an emergent property of a particular partition of neuronal states, over progressively larger scales. As such, it offers a way of linking dynamics on directed graphs to the phenomenology of intrinsic brain networks.

# Introduction

A persistent theme in systems neuroscience, especially neuroimaging, is the search for principles that underlie the functional anatomy of distributed neuronal processes. These principles are usually articulated in terms of functional segregation (or differentiation) and integration – that inherit from centuries of neuroanatomical, neurophysiological, and neuropsychological study (Zeki & Shipp, 1988). In recent thinking about functional integration, people have turned to formal accounts of (predictive) processing in the brain; e.g., (A M Bastos et al., 2012; Keller & Mrsic-Flogel, 2018; Parr & Friston, 2018; Rao & Ballard, 1999; Spratling, 2008) to understand the nature of (neuronal) message passing on graphs: where edges correspond to connectivity and nodes correspond to neuronal populations. Crucially, this characterisation rests upon the asymmetric and directed connectivity that defines cortical and subcortical hierarchies; e.g., (A M Bastos et al., 2012; Crick & Koch, 1998; Felleman & Van Essen, 1991; K. J. Friston, Parr, & de Vries, 2017; Keller & Mrsic-Flogel, 2018; Markov et al., 2013; Mesulam, 1998; Stachenfeld, Botvinick, & Gershman, 2017; Zeki & Shipp, 1988). Usually, these asymmetries are expressed in terms of things like laminar specificity that distinguish between forward and backward connections (Buffalo, Fries, Landman, Buschman, & Desimone, 2011; Grossberg, 2007; Haeusler & Maass, 2007; Hilgetag, O'Neill, & Young, 2000; Thomson & Bannister, 2003; Trojanowski & Jacobson, 1977). More recently, asymmetries in spectral content have become an emerging theme (Arnal & Giraud, 2012; A. M. Bastos et al., 2015; Buffalo et al., 2011; Giraud & Poeppel, 2012; Hovsepyan, Olasagasti, & Giraud, 2018; Self, van Kerkoerle, Goebel, & Roelfsema, 2019; Singer, Sejnowski, & Rakic, 2019; van Kerkoerle et al., 2014).

In contrast, analyses of functional connectivity have focused on distributed patterns of coherent fluctuations in neuronal activity and phenomenological descriptions of the implicit dynamics (Bassett & Sporns, 2017; Biswal, Van Kylen, & Hyde, 1997; Bullmore & Sporns, 2009; Gilson, Moreno-Bote, Ponce-Alvarez, Ritter,

& Deco, 2016; Gu et al., 2018; Lynall et al., 2010; van den Heuvel & Sporns, 2013). This phenomenology ranges from intrinsic brain networks – that are conserved over subjects in resting state functional magnetic resonance imaging – to the dependence of neuronal dynamics on cortical excitability (Freyer, Roberts, Ritter, & Breakspear, 2012; Roy et al., 2014). The principles that are brought to bear on this kind of characterisation could be seen as ascribing neuronal dynamics to various universality classes, such as self-organised criticality (Bak, Tang, & Wiesenfeld, 1988; Michael Breakspear, Heitmann, & Daffertshofer, 2010; Cocchi, Gollo, Zalesky, & Breakspear, 2017; Deco & Jirsa, 2012; K. J. Friston, Kahan, Razi, Stephan, & Sporns, 2014; Haimovici, Tagliazucchi, Balenzuela, & Chialvo, 2013; Kitzbichler, Smith, Christensen, & Bullmore, 2009; Shin & Kim, 2006). This dual pronged approach to functional integration invites an obvious question – is there a way of linking the two?

Practically, the study of context-sensitive, directed coupling between the nodes of neuronal networks calls for an estimate of effective connectivity, under some model of how measured brain signals are generated. One then has to resolve the ill-posed problem of recovering the underlying (connectivity) parameters of the model; usually using Bayesian inference. The best example here is dynamic causal modelling (K J Friston, Harrison, & Penny, 2003). The complementary approach – based upon functional connectivity – borrows ideas from network science and graph theory. This entails specifying an adjacency matrix, usually formed by thresholding a functional connectivity matrix summarising dependencies among nodes, where the nodes are generally defined in terms of some parcellation scheme (Bassett & Sporns, 2017; Bullmore & Sporns, 2009).

In what follows, we will consider a particular parcellation scheme based upon effective connectivity and ask whether it leads to the same phenomenology seen in network neuroscience. In doing so, we can, in principle, explain and quantify the emergence of large-scale intrinsic brain networks and their characteristic dynamics. A crucial aspect of the particular parcellation or partition – employed in this work – means that it can be applied recursively in the spirit of the renormalisation group (Schwabl, 2002). This means that there is a formal way of quantifying the dynamics at various spatiotemporal scales. Our hypothesis was that the spatiotemporal dynamics of larger scales would evince both the functional anatomy of intrinsic brain networks – and the emergence of (self-organised) criticality – as assessed in terms of dynamical instability.

Although this work is framed as addressing issues in network neuroscience (Bassett & Sporns, 2017), it was originally conceived as a parcellation scheme for multiscale analyses of neuroimaging timeseries. In other words, it was intended as a first principle approach to dimension reduction and decomposition, as a prelude for subsequent graph theoretic or dynamic causal modelling. However, the theoretical foundations

– and uniqueness of the partition – proved too involved to support a simple and practical procedure. Instead, what follows is offered as a case study of emergence in coupled dynamical systems, using the brain as a paradigm example.

This paper comprises five sections. In the first, we review the notion of Markov blankets and how recursive applications of a partition or parcellation of states into Markov blankets allows one to express dynamics at increasing scales. We will use the notion of the renormalisation group (**RG**) to motivate this recursive parcellation because there are some formal constructs (in terms of **RG** scaling) that furnish an insight into how dynamics change as we move from one scale to the next. The second section describes a simple (dynamic causal modelling) analysis of directed effective connectivity at the lowest spatial scale, as summarised with a Jacobian. This plays the role of a directed adjacency matrix, which is all that is needed for successive renormalisation to higher scales. The renormalisation group is illustrated with an exemplar dataset, to show what the ensuing parcellation scheme looks like. This section concludes with a brief consideration of sparse coupling at the lowest scale, in terms of excitatory and inhibitory connections. The subsequent sections consider dynamics at different scales of parcellation, in terms of intrinsic (within parcel) and extrinsic (between parcel) connectivity. Our focus here is on the progressive slowing of intrinsic dynamics as we move from one scale to the next – a slowing that organises the dynamics at larger (higher) scales towards critical regimes of instability and slowly fluctuating dynamical modes. The third section illustrates the emergence of autonomous dynamics – in terms of characteristic frequencies associated with intrinsic connectivity – and in terms of positive Lyapunov exponents that speak to transcritical bifurcations at, and only at, larger scales. The fourth section focuses on extrinsic connectivity and the coupling between (complex) modes or patterns of activity and how this relates to functional connectivity and intrinsic brain networks (Fox et al., 2005). The final section reviews the dynamical phenomenology at hand from the point of view of statistical physics, with a special focus on dissipative dynamics and detailed balance at non-equilibrium steady-state. We conclude with a brief discussion and qualification of this particular (sic) approach to functional integration.

## Markov blankets and the renormalisation group

The last section concluded with reference to a particular partition. The use of the word "particular" has a *double entendre* here. It is predicated on a more fundamental (or perhaps foundational) analysis of coupled dynamical systems that consider the emergence of "particles". Full details of this treatment can be found in (Karl Friston, 2019). From the current perspective, we just need to know how to define Markov blankets (Clark, 2017; Kirchhoff, Parr, Palacios, Friston, & Kiverstein, 2018; J Pearl, 1988; Pellet & Elisseeff, 2008) and how Markov blankets engender particles and particular partitions (Karl Friston, 2019).

In brief, a Markov blanket allows one to distinguish a collection of *vector states* (hereafter, simply states) that belong to a particle from states that do not. This provides an operational definition of a particle which, in the present setting, can be regarded as a region of interest or *parcel* of brain states. This means that a particular partition becomes a parcellation scheme, in terms of functional anatomy. The particular partition refers to a partition of a (potentially large) set of states into a smaller number of particles, where each particle is distinguished from other particles, in virtue of possessing a Markov blanket. A Markov blanket is simply a set of states that separates or insulates – in a statistical sense – states that are internal to the blanket and states that are on the outside; namely, external states. Technically, this means that internal states are conditionally independent of external states, when conditioned upon their blanket states (Judea Pearl, 2009).

In a particular partition, all external states are assigned to particles – to create an ensemble of particles that are constituted by their blanket states and the internal states within or beneath the blanket. The crucial aspect of this partition is that we only need the blanket states to understand coupling between particles. This follows from the conditional independence between internal and external states, where the external states 'that matter' are the blanket states of other particles. In short, the particular partition is a principled way of dividing states into particles or parcels that is defined in terms of statistical dependencies or coupling among states. In more complete treatments, one can divide the blanket states into active states and sensory states, according to the following rules: sensory states are not influenced by internal states, while active states are not influenced by external states. Indeed, it is the absence of these influences that enables us to identify the Markov blanket of any given set of internal states. Please see the appendix for a formal definition of Markov blankets in this dynamical context.

As noted above, we are dealing with vector states (not scalar variables). So, what is a vector state? A vector state is the multidimensional state of a particle, for example, the principal eigenstates of its Markov blanket. However, we have just said that a particle arises from a partition of states – and now we are saying that a state is an eigenstate (i.e., a linear mixture) of the blanket states of a particle. So, is a particle a collection of states or is a state the attribute of a particle (i.e., its blanket states)? The answer is both because we have particles at multiple levels.

This is where the renormalisation group comes in, via a recursive application of the particular partition. In other words, if we start with some states at any level, we can partition these states into a set of particles – based upon how the states are coupled to each other. We can then take the principal eigenstates of each particle's blanket states to form new states at the scale above – and start again. This recursive application of a *grouping* or partition operator (**G**) – followed by a dimension reduction (**R**) – leads to the renormalisation group based upon two operators, **R** and **G**. In theoretical physics, the renormalization group (**RG**) refers to a transformation that characterises a system when measured at different scales (Cardy, 2015; Schwabl, 2002). A working definition of renormalization rests on three things (Lin, Tegmark, & Rolnick, 2017): vectors of random variables, a coarse-graining operation and a requirement that the operation does not change the functional form of the Lagrangian to within a multiplicative constant. For example, under a transformation of position and velocity variables $x$ and $\dot{x}$ given by $x \to ax$ and $\dot{x} \to b\dot{x}$, the corresponding Lagrangian $\lambda$ transforms (if scale-invariant) according to $\lambda(x, \dot{x}) \to \lambda(ax, b\dot{x}) = c\lambda(x, \dot{x})$, where $a, b$ and $c$ are constants (Landau, 1976). Equivalently, a scale invariant system's equation of motion must remain *perfectly* unchanged under the re-scaling operation. This can readily be seen by applying the Euler-Lagrange equation to the scaled Lagrangian:

$$\frac{d}{dt}\left[\frac{\partial(c\lambda)}{\partial\dot{x}}\right] = \frac{\partial(c\lambda)}{\partial x} \Rightarrow c\frac{d}{dt}\left[\frac{\partial(\lambda)}{\partial\dot{x}}\right] = c\frac{\partial(\lambda)}{\partial x} \tag{1}$$

Here, the re-scaling constant $c$ cancels, leaving the original equation of motion[1].

In our case, the random variables are states; the coarse graining operation corresponds to the grouping into a particular partition (**G**) and a dimension reduction (**R**) inherent in retaining the principal eigenstates of particular blanket states. The dimension reduction operator (**R**) has two parts. First, we can eliminate the internal states because they do not contribute to coupling between particles. Second, we can eliminate the eigenstates that dissipate very quickly; namely, those with large negative eigenvalues. These are the fast or stable modes of a dynamical system (Carr, 1981; Haken, 1983). This leaves us with the slow unstable eigenstates picked out by the dimension reduction, which we can now see as an adiabatic approximation[2].

---

[1] In what follows, instead of dealing with real positions and velocities, we will deal with complex variables that have real and complex parts.
[2] In quantum mechanics, the adiabatic approximation refers to those solutions to the Schrödinger equation that make use of a time-scale separation between fast and slow degrees of freedom.

Formally, we can express the coarse graining or *blocking* transformation $\mathbf{R} \circ \mathbf{G}$ as a composition of a particular partition and adiabatic reduction applied to any random dynamical system (at scale $i$) that can be characterised as coupled subsets of states. The $n$-th subset $x_n^{(i)} \subset x^{(i)}$ constitutes the vector state of a *particle*, subject to random fluctuations, $\omega_n^{(i)}$:

$$\dot{x}_n^{(i)} = \sum_m \lambda_{nm}^{(i)} x_m^{(i)} + \omega_n^{(i)} \Rightarrow J(x_n^{(i)}, x_m^{(i)}) \triangleq \frac{\partial \dot{x}_n^{(i)}}{\partial x_m^{(i)}} = \lambda_{nm}^{(i)} \tag{2}$$

These equations of motion for the states of the $n$-th particle comprise intrinsic and extrinsic components, determined by the states of the particle in question and other particles, respectively. In this form, the diagonal elements of the Jacobian or coupling matrix, $\lambda_{nn}^{(i)} \in \mathbb{C}$, determine the frequency and decay of oscillatory responses to extrinsic perturbations and random fluctuations. The grouping operator ($\mathbf{G}$) groups states into particles, where particles comprise blanket and internal states: $\pi_j^{(i)} = \{b_j^{(i)}, \mu_j^{(i)}\}$. The blocking transformation ($\mathbf{R}$) then reduces the number of states, by eliminating internal states at the lower level and retaining slow eigenstates using the principal eigenvectors $\xi_n^{(i)} = eig(J(b_n^{(i)}, b_n^{(i)}))$ of the Jacobian of blanket states $b_n^{(i)}$. These eigenstates then become the vector states at the next scale:

$$\left\{ x_n^{(i)} \right\} = \mathbf{R} \circ \mathbf{G} \circ \left\{ x_n^{(i-1)} \right\}$$

$$\{\lambda_{nm}^{(i)}\} = \beta(\{\lambda_{nm}^{(i-1)}\})$$

$$\{x_n^{(i)}\} \xrightarrow{\ \mathbf{G}\ } \{\pi_j^{(i)}\} : \pi_j^{(i)} = \{b_j^{(i)}, \mu_j^{(i)}\}$$

$$\{b_n^{(i)}\} \xrightarrow{\ \mathbf{R}\ } \{x_n^{(i+1)}\} = \{\xi_n^{(i)-} b_n^{(i)}\} : \xi_n^{(i)} = eig(J(b_n^{(i)}, b_n^{(i)}))$$

$$\{\lambda_{nm}^{(i)}\} \xrightarrow{\ \beta\ } \{\lambda_{nm}^{(i+1)}\} = \{\xi_n^{(i)-} J(b_n^{(i)}, b_m^{(i)}) \xi_m^{(i)}\} \tag{3}$$

Here, the parameters of the Lagrangian are taken to be the coupling parameters $\lambda_{nm}^{(i)}$, whose changes are implemented by a *beta function* that is said to induce a renormalization group flow (or $\mathbf{RG}$ flow). The key aspect of this flow rests upon the adiabatic reduction, which renders the dynamics progressively slower at successive macroscopic scales. This follows because, by construction, only slow eigenstates are retained, where the intrinsic coupling among these eigenstates is a diagonal matrix of (negative) eigenvalues, which determine how quickly the eigenstates decay:

$$E[Re(\lambda_{nn}^{(i)})] \leq E[Re(\lambda_{nn}^{(i+1)})] \dots \leq 0 \tag{4}$$

The **RG** flow speaks to a progressive move from dynamics with high amplitude, fast fluctuations (e.g., quantum mechanics) through to deterministic systems that are dominated by slow dynamics (e.g., classical mechanics). In deterministic systems, the real parts of $\lambda_{nn}^{(i)}$ play the role of *Lyapunov exponents* (c.f., critical exponents), which quantify the rate of separation of infinitesimally close trajectories (Lyapunov & Fuller, 1992; Pyragas, 1997). This suggests that as we move from one scale to the next, there is a concomitant increase in the tendency to critical slowing and dynamic itinerancy (Cessac, Blanchard, & Krüger, 2001; Pavlos, Karakatsanis, & Xenakis, 2012).

In this (**RG**) setting, a *relevant* variable is said to describe the macroscopic behaviour of the system. From our perspective, the relevant variables in question correspond to the slow eigenstates. In short, we can reduce many states to a small number of eigenstates that summarise the dynamics 'that matter'. These eigenstates are the relevant variables that underwrite critical slowing. Figures 1 and 2 provide a graphical illustration of this recursive partitioning and reduction based upon an adiabatic approximation (i.e., eliminating fast eigenstates and approximating dynamics with the remaining slow eigenstates). This adiabatic reduction is commonplace in physics, where it plays a central role in synergetics through the enslaving principle (Haken, 1983) and, in a related form, in the centre manifold theorem (Carr, 1981).

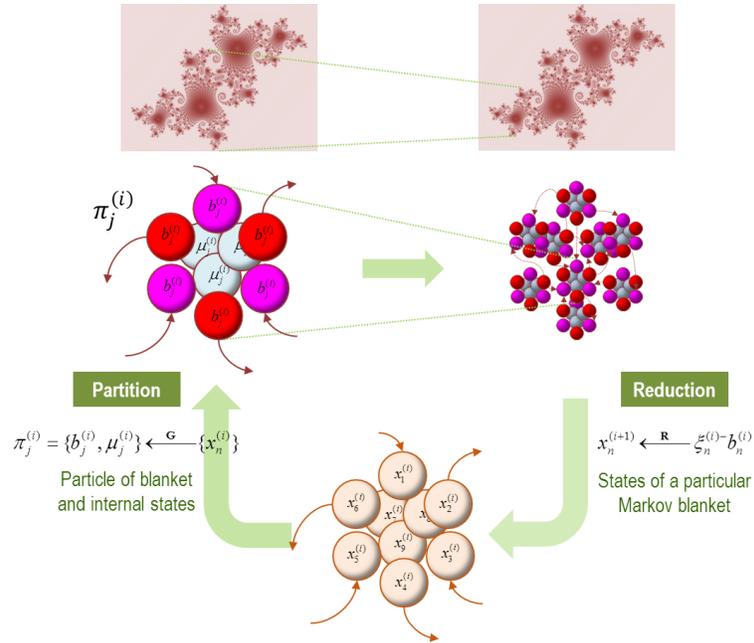

**Figure 1**

**Blankets of blankets**. This schematic illustrates the recursive procedure by which successively larger scale (and slower) dynamics arise from subordinate levels. At the bottom of the figure (lower panel), we start with an ensemble of vector states (here nine). The conditional dependencies among these vector states (i.e., eigenstates) define a particular partition into particles (upper panels). Crucially, this partition equips each particle with a bipartition into blanket and internal states, where blanket states comprise active (red) and sensory states (magenta). The behaviour of each particle can now be summarised in terms of (slow) eigenstates or mixtures of its blanket states to produce states at the next level or scale. These constitute an ensemble of vector states and the process starts again. Formally, one can understand this in terms of coarse graining the dynamics of a system via two operators. The first uses the particular partition to *group* subsets of states (**G**), while the second uses the eigenstates of the resulting blanket states to *reduce* dimensionality (**R**). The upper panels illustrate the bipartition for a single particle (left panel) and an ensemble of particles, i.e., the particular partition *per se* (right panel). The insets on top illustrate the implicit self-similarity of particular dependencies pictorially, in moving from one scale to the next. Please see the main text for a definition of the variables used in this figure.

We now have at hand a principled procedure to repeatedly coarse-grain a system of loosely coupled particles (e.g., nonlinear neuronal oscillators) at successively larger spatiotemporal scales. One can see that, by construction, as we ascend scales, things will get larger and slower. It is this progressive slowing towards criticality that is the primary focus of the examples pursued below. However, before we can apply the particular partition to some empirical data, we first need to quantify the coupling among particles at a suitably fine or small scale. Having characterised this coupling in terms of some dynamical system or state space model, we can then use the Jacobian to identify a particular partition, compute the Jacobian of the blanket states and then take the ensuing eigenstates to the next level, as described above. This furnishes a description of dynamics in terms of the intrinsic (within particle) coupling (i.e., eigenvalues) of any particle $\lambda_{nn}^{(i)}$ and their extrinsic (between particle) coupling $\lambda_{nm}^{(i)}$. We will unpack the meaning of these terms later using numerical examples and analysis. At present, we will focus on estimating the coupling among a large number of particles at the smallest scale.

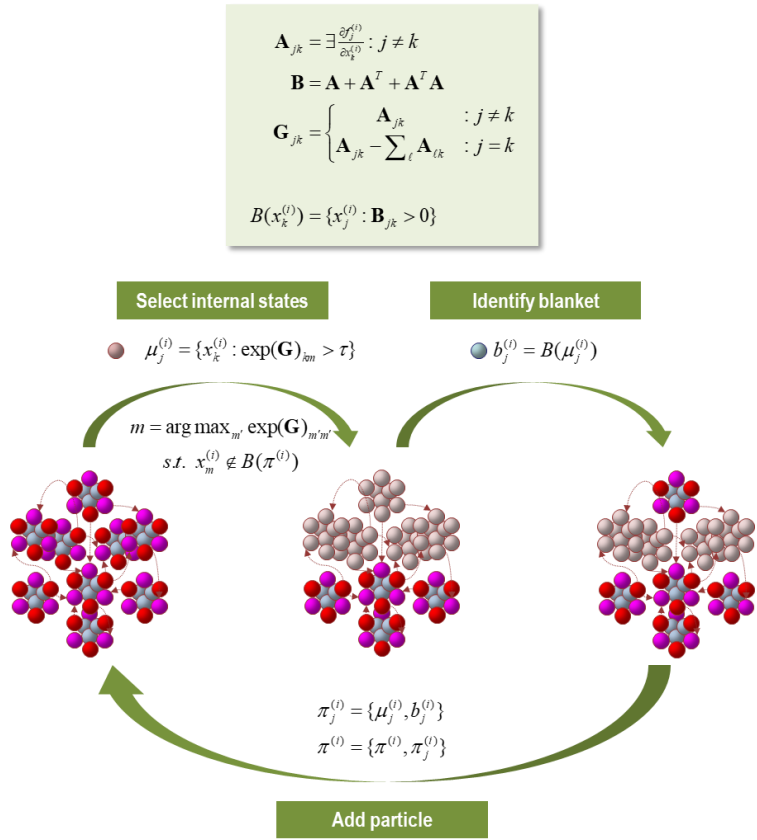

**The particular partition**. This schematic illustrates a partition of eigenstates (small coloured balls) into particles (comprising nine vectors), where each particle has six blanket states (red and magenta for active and sensory states respectively) and three internal states (cyan). The upper panel summarises the operators used to create a particular partition. We start by forming an adjacency matrix that characterises the coupling between different vectors states. This is based upon the Jacobian and implicitly the flow of vector states. The resulting adjacency matrix defines a Markov blanket forming matrix (**B**), which identifies the children, parents, and parents of the children. The same adjacency matrix is used to form a graph Laplacian (**G**) that is used to define neighbouring (i.e., coupled) internal states. One first identifies a set of internal states using the graph Laplacian. Here, the $j$-th subset of internal states at level $i$ are chosen, based upon dense coupling with the vector state with the largest graph Laplacian. Coupled internal states are then selected from the columns of the graph Laplacian that exceed some threshold. In practice, the examples used later specify the number of internal states desired for each level of the hierarchical decomposition. Having identified a new set of internal states (that are not members of any particle that has been identified so far) its Markov blanket is recovered using the Markov blanket forming matrix. The internal and blanket states then constitute a new particle, which is added to the list of particles identified. This procedure is repeated until all vector states have been accounted for. Usually, towards the end of this procedure, candidate internal states are exhausted because all remaining unassigned vector states belong to the Markov blanket of the particles identified previously. In this instance, the next particle can be an active or sensory state, depending upon whether there is a subset (of active states) that is not influenced by another. In the example here, we have already identified four particles and the procedure adds a fifth (top) particle to the list of particles; thereby accounting for nine of the remaining vector states.

## Starting from the bottom

To use the machinery of Markov blankets, in the setting of loosely coupled dynamical systems, we need to specify the coupling among vector states (that we can associate with the eigenstates of the smallest particles under consideration). To do this, one can use a simplified form of dynamic causal modelling that can be applied to hundreds or thousands of neuronal states. This is easier than it might sound, provided one commits to low (first) order approximations; e.g., (Frassle et al., 2017). Consider the state space model describing the coupling among a large number of states, where the flow is subject to random fluctuations (dropping superscripts for clarity):

$$\dot{x} = f(x) + \omega_x$$
$$y = k * x + \omega_y$$

(5)

Notice that we have introduced a convolution operator that converts latent (neuronal) states to some observable measurement (e.g., haemodynamic signals from functional magnetic resonance imaging). Here, $y(t)$ is a linear convolution (with kernel $k$) of some states $x(t)$ subject to observation and system noise, respectively. We have also assumed that there is an observation for each *relevant* state. Linearising this state space model, where $J = \partial_x f(x)$ and † denotes conjugate transpose, we have:

$$\begin{matrix} Dx = xJ^\dagger + \omega_x \\ y = Kx + \omega_y \end{matrix} \Rightarrow \begin{Bmatrix} KDx = KxJ^\dagger + K\omega_x \\ Dy = KDx + D\omega_y \end{Bmatrix} \Rightarrow \begin{cases} Dy = yJ^\dagger + \omega \\ \omega = K\omega_x + D\omega_y - \omega_y J^\dagger \end{cases}$$

(6)

Here, the states have been arranged into a matrix, with one row for each point in time and a column for each dimension. This means we can replace the derivative and convolution operators in equation (5) with the matrix operators in equation (6) that commute[3], i.e., $KD = DK$. In turn, this means we can approximate the system with a general linear model:

$$Dy = yJ^\dagger + \omega$$
$$\text{cov}(\omega) = \gamma_1 KK^\dagger + \gamma_2 DD^\dagger + \gamma_3 I \tag{7}$$

This approximation assumes that $J^\dagger \ J \propto I$. This assumption is licensed by the fact that the Jacobian of *relevant* states will be dominated by large negative leading diagonals (that underwrite the stability of each state). Equation (7) is a straightforward general linear model, with random fluctuations that have distinct covariance components, which depends upon the form of the (e.g., haemodynamic) convolution kernel and the amplitude of state and observation noise. If $K$ is specified in terms of the basis set of convolution kernels, then the covariance components of the linearised system can be expressed as:

$$K = \sum_k \kappa_k K_k \Rightarrow$$
$$KK^\dagger = \sum_{ij} \kappa_i \kappa_j K_i K_j^\dagger \tag{8}$$

Such that $\kappa_i \kappa_j$ replaces the hyperparameter $\gamma_1$ above.

This linearized system can now be solved using standard (Variational Laplace) schemes for parametric empirical Bayesian (PEB) models, to provide (approximate) Gaussian posteriors over the unknown elements of the Jacobian – and the unknown covariance parameters encoding the amplitude of various random effects (K Friston, Mattout, Trujillo-Barreto, Ashburner, & Penny, 2007). This Bayesian model inversion requires priors on the parameters and hyperparameters (i.e., covariance component parameters), specified as Gaussian shrinkage priors. For nonnegative hyperparameters, Gaussian shrinkage priors are generally applied to log transformed hyperparameters (i.e., a lognormal prior over nonnegative scale parameters).

Equipped with posterior densities over the coupling parameters – or elements of the Jacobian – we can now use Bayesian model reduction to eliminate redundant parameters (K. J. Friston et al., 2016); namely, parameters whose shrinkage to zero increases model evidence by removing redundancy or complexity. As described elsewhere (K Friston & Penny, 2011), this can be done very efficiently, because we know the form of the posteriors, before and after reducing the model. Furthermore, we can apply other prior constraints to eliminate redundant coupling parameters.

---

[3] This is due to the linearity of the convolution operator and is true whether the temporal derivative is in matrix form or not. Intuitively, a linear combination of velocities is equivalent to the rate of change of a linear combination of positions.

In the examples below, we performed Bayesian model reduction to enforce reciprocal coupling among states, given that extrinsic connections in the brain are almost universally recurrent (Felleman & Van Essen, 1991; Markov et al., 2013). This was implemented by combining the changes in variational free energy – or log model evidence – when removing connections between two states in both directions. If model evidence increased by three natural units (i.e., a log odds ratio of exp(3):1 or 20:1), both connections were removed but not otherwise. In addition, we precluded long range coupling (beyond 32 mm) and used Bayesian model reduction to identify the most likely upper bound on the spatial reach of coupling between non-homologous particles (i.e., particles that did not occupy homologous positions in each hemisphere). These empirical connectivity priors were based upon a body of empirical work, suggesting that the density of axonal projections – from one area to another – declines exponentially as a function of anatomical separation (Finlay, 2016; Horvát et al., 2016; Wang & Kennedy, 2016). We will later examine the evidence for this kind of distance rule, based upon coupling among particles at the smallest scale.

## Functional parcellation

Computationally, the benefit of linearising the system in this way means that one can evaluate the posterior coupling parameters or elements of the Jacobian region by region; c.f., (Frassle et al., 2017). This means that, provided one is prepared to wait long enough, one can invert large systems with thousands of regions or parcels. On a personal computer, it takes about an hour to evaluate the Jacobian for coupling among 1024 states. To illustrate the renormalisation group procedure practically, we applied it to the fMRI timeseries from a single subject. These timeseries are the same data used to illustrate previous developments in dynamic causal modelling[4].

---

[4] Timeseries data were acquired from a normal subject at 2 Tesla using a Magnetom VISION (Siemens, Erlangen) whole body MRI system. Contiguous multi-slice images were acquired with a gradient echo-planar sequence (TE = 40ms; TR = 3.22 seconds; matrix size = 64x64x32, voxel size 3x3x3mm). Four consecutive hundred-scan sessions were acquired, comprising a sequence of 10-scan blocks under five conditions. The first was a dummy condition to allow for magnetic saturation effects. In the second, *Fixation*, the subject viewed a fixation point at the centre of the screen. In an *Attention* condition, the subject viewed 250 dots moving radially from the centre at 4.7 degrees per second and was asked to detect changes in radial velocity. In *No attention*, the subject was asked to look at the moving dots. In last condition, subject viewed stationary dots. The order of the conditions alternated between *Fixation* and photic stimulation. The subject fixated the centre of the screen in all conditions. No overt response was required in

In the exemplar analyses below, we started at a scale where each particle can be plausibly summarised with a single state. This single state was the principal eigenstate following a principal components analysis of voxels that lay within about 4 mm of each other. This can be thought of as reducing the dynamics to a single mode of the Markov blanket of this small collection of voxels. Practically, this simply involved taking all voxels within a fixed radius of the voxel showing the largest variance, performing a singular value decomposition, and recording the first eigenvariate. These voxels were then removed, and the procedure repeated until the entire multivariate timeseries was reduced to 1024 eigenstates, where each eigenstate corresponds to a simple particle. See Figure 3. Clearly, we could have summarised the dynamics of each collection of voxels with two or more eigenstates; however, for simplicity we will assume that the eigenstate with the greatest variance is a sufficient summary of the slow, non-dissipative, dynamics of this smallest scale. Interestingly, this variance is proportional to the characteristic time constant of systemic dynamics; namely, the negative inverse of the eigenvalues of the underlying Jacobian (see final section). In other words, as the (negative) principal eigenvalue of effective connectivity approaches zero from below, the principal eigenvalue of functional connectivity (i.e. variance) increases: see equation (9) in (K. J. Friston, Kahan, et al., 2014).

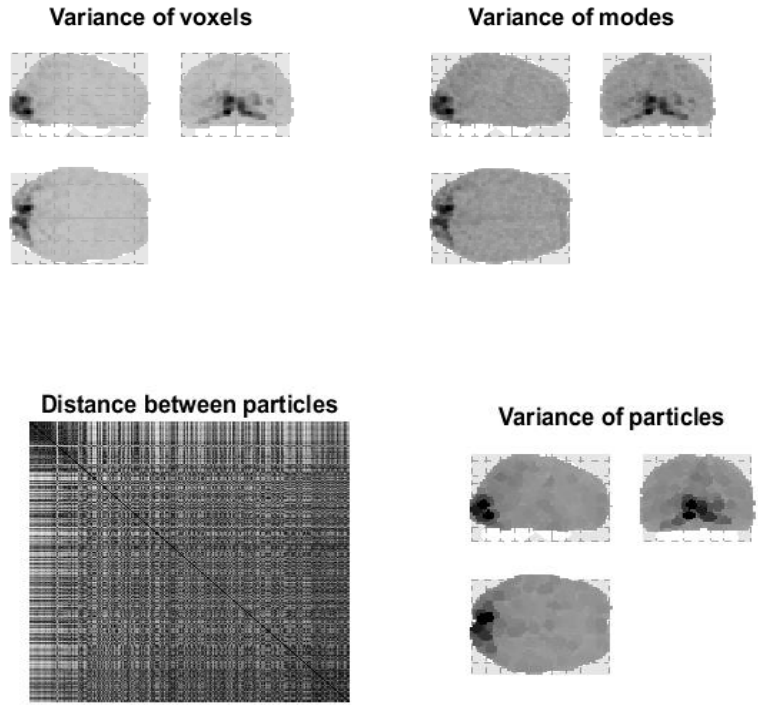

**Distributed variance**: this figure illustrates the variance explained by particles at the first level. The **upper panel** is a maximum intensity projection of the variance of the fMRI timeseries, for a single subject over 360 scans (with a repetition time of 3.22 seconds) in voxel space. One can see that visual (i.e., striate) and extrastriate regions have been preferentially engaged; however, there is distributed activity throughout the brain. The upper right panel shows the corresponding variance in terms of the eigenmodes of 1024 particles. As in subsequent figures, these projections involve weighting the absolute value of each eigenmode by the quantity in question; here, the variance. This maximum intensity projections shows that the particles furnish a reasonably faithful summary of voxels-specific variance. The **lower right** panel shows the same variance assigned to the spatial support of each eigenmode, to illustrate the coarse graining when assembling particles from voxels. These characterisations of fluctuations over time have been framed in terms of variance. We will see later that variance can be interpreted as a dissipative time constant. In other words, in this example, visual areas show the least dissipation, with dynamics that decay slowly. The **lower left panel** shows the Euclidean distance between the centres of pairs of particles. The centre of each article was defined as the expected anatomical location, where the probability density over location was taken to be a softmax function of the absolute value of the eigenmode over voxels. In this and subsequent figures, Euclidean distances are evaluated after projecting centres across the sagittal plane, i.e., superimposing homologous particles in the right and left hemispheres. This means a particle in the left hemisphere is 'close to' a homologous particle in the right hemisphere.

Following Bayesian model reduction (see Figure 4), we now have a sparse Jacobian or directed, weighted adjacency matrix describing the dynamical coupling between univariate states of 1024 particles (see Figure 4). This Jacobian can now be subject to a particular partition to identify the most connected internal states and their Markov blanket – following the procedures summarised in Figure 2. This grouping process (i.e., the **G** operator) was repeated until all 1024 states are accounted for: in this example, grouping 1024 states into 57 particles. For simplicity, and consistency with the first level, each particle was assigned a single internal state. The ensuing partition was then subject to an adiabatic reduction (i.e., the **R** operator) by taking the eigenvectors of the Jacobian describing the intrinsic dynamics of each particle's blanket states.

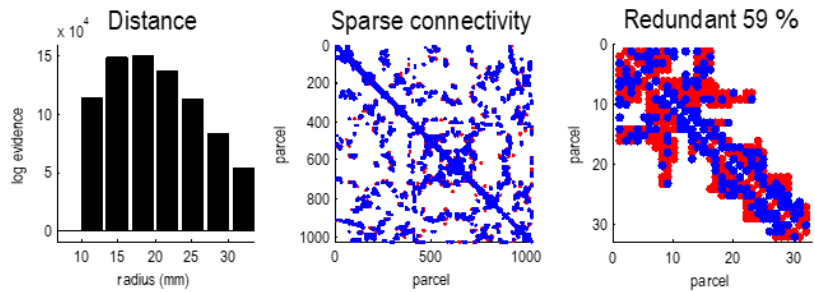

**Figure 4**

**Sparse connectivity**: this figure illustrates the sparsity of effective connectivity using Bayesian model reduction. The **left panel** shows the log evidence for a series of models that precluded connections beyond a certain distance or radius. This log evidence has been normalised to the log evidence of the model with the least marginal likelihood (i.e., coupling over less than 10 mm). These results show that a model with local connectivity (about 18 mm) has the greatest evidence. In other words, effective connections beyond this distance are redundant, in the sense that they add more complexity to log evidence that is licensed by an increase in accuracy. The **middle panel** shows the ensuing sparse coupling (within the upper bound of 32 mm) as an adjacency matrix, where particles have been ordered using a nearest neighbour scheme in voxel space. The blue dots indicate connections that have been removed by Bayesian model reduction. In this instance, nearly 60% of estimated connections were redundant. The **right panel** zooms in on the first 32 particles, to show some local connections that were retained (red) or removed (blue).

The eight principal eigenstates were retained if their eigenvalues were less than one. This is the adiabatic approximation that dispenses with modes that dissipate quickly, here, within a second. This reduces the intrinsic coupling to a diagonal matrix $\lambda_{nn}^{(i)}$, corresponding to the eigenvalues of the intrinsic Jacobian $\partial_{x_n} f_n^{(i)}$. The extrinsic coupling $\lambda_{nm}^{(i)}$ contain complex elements that couple the eigenstates of one particle to the eigenstates of another. We will return to how these Jacobians manifest in terms of connectivity later.

In short, we now have a summary of dynamics at the scale above in terms of the eigenstates of a particle that, by construction, have been orthogonalised. These constitute the vector states for the next application of the **RG** operator to produce a description of dynamics at subsequent scales. See Figure 5 through to Figure 8. These examples show that by the fourth scale we have reduced the dynamics to a single particle, shown in a maximum intensity projection format in Figure 8. We can project particles onto anatomical space because each state that constitutes a particle at any scale is a mixture of states that, ultimately, can be associated with a particular location in voxel space. In other words, particles inherit a spatial location from the scale below, enabling one to visualise (and quantify) the spatial scale of particles at successively higher scales. We will refer to this characterisation of an eigenstate as an *eigenmode*; namely, a pattern in voxel space whose amplitude is determined by the corresponding eigenstate. One can express the eigenmodes in terms of the eigenvectors at each scale as follows:

$$v_{n_j}^{(i)} = \xi^{(1)} \xi^{(2)} \dots \xi_{n_j}^{(i)}, \quad \xi^{(i)} = \begin{bmatrix} \xi_1^{(i)} & & \\ & \ddots & \\ & & \xi_N^{(i)} \end{bmatrix} \tag{9}$$

This gives the eigenmode of the *j*-th eigenstate of the *n*-th particle at the *i*-th scale.

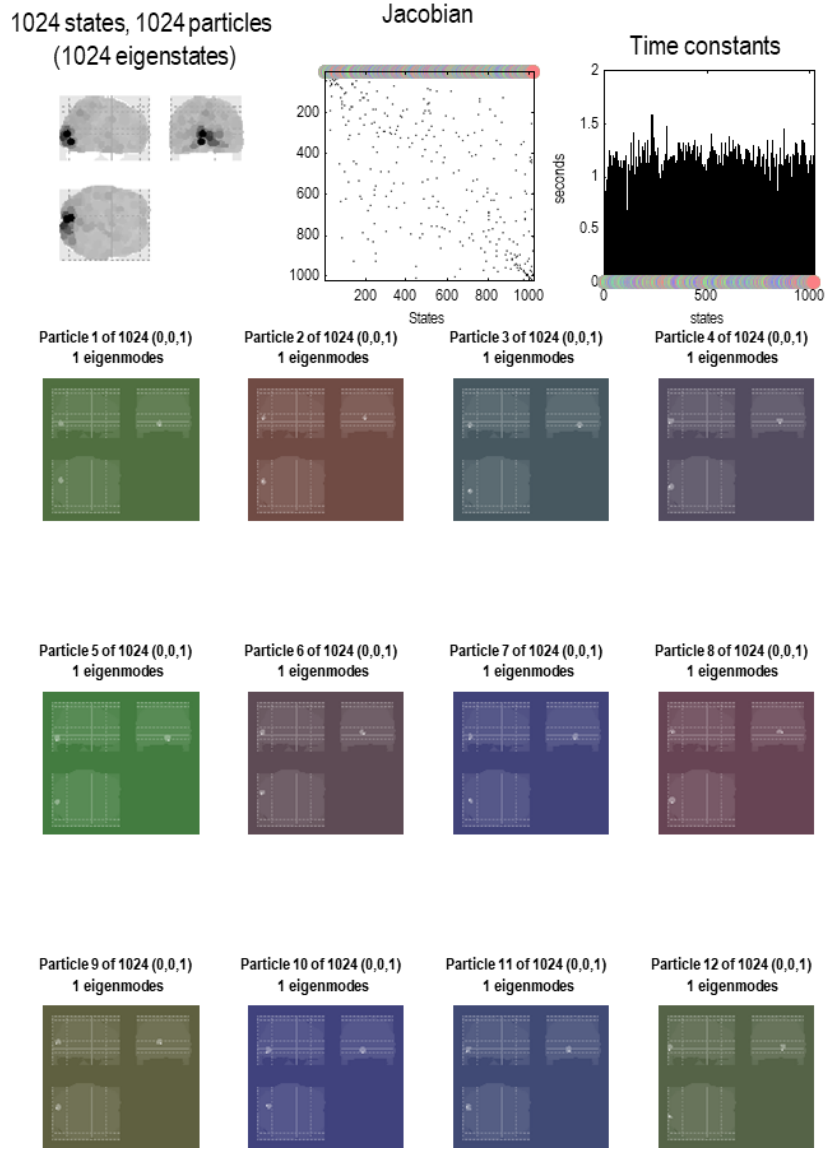

**Figure 5**

**Brain particles**: This figure illustrates the partition of states at the first level. The format of this figure is replicated in subsequent figures that detail a particular decomposition at increasing scales. The **upper left panel** shows all the constituent particles as a maximum intensity projection, where the spatial support of each particle has been colour-coded according to the variance explained by its eigenmode. One can see that nearly the entire brain volume has been effectively tiled by 1024 particles. The **upper middle panel** shows the corresponding adjacency matrix or coupling among particles. The coloured circles encode the identity of each particle. In this instance, the particles have been arranged in order of descending dissipation (i.e., the real part of the eigenvalue of each particle's Jacobian). The **upper right panel** shows these eigenvalues above the corresponding particle (encoded by coloured dots) in terms of rate constants (i.e., the negative inverse of the eigenvalues). The **remaining panels** show the first 12 particles as maximum intensity projections. The colour of the background corresponds to the colour that designates each particle. In this first level, each particle has a single eigenstate. The numbers in brackets above each maximum intensity projection correspond to the number of internal, active, and sensory states, respectively, where the active and sensory states comprise blanket states. At this lowest level, every eigenstate is a sensory state because it can influence – and be influenced by – the eigenstates of other particles. At this scale, one can see the particles are small, with a standard deviation of about 3 mm (based on the softmax function of the absolute value of each particle's eigenmodes). Here, the characteristic time constants of these particles are about one second. This should be compared with the equivalent distribution in the upper right panel of the next figure.

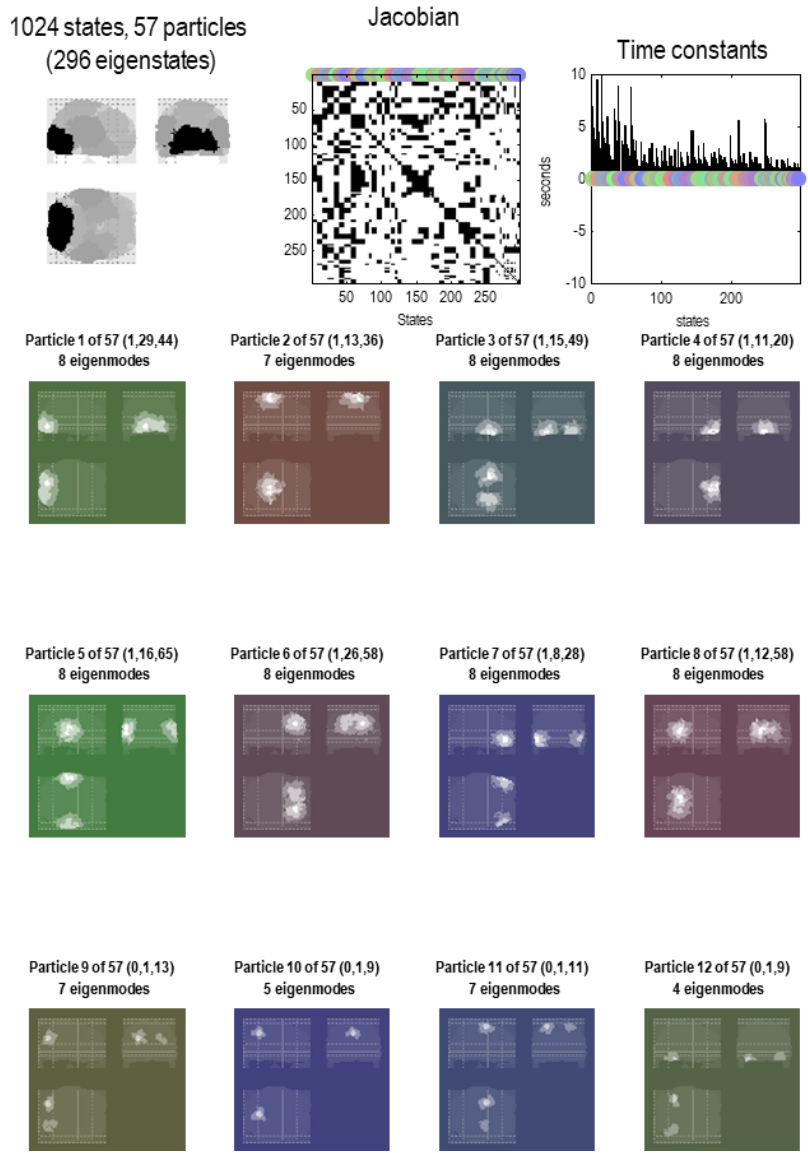

**Figure 6**

**Particular partition at the second scale**: this figure uses the same format as the previous figure; however, here, we are looking at particles at the next scale. In other words, aggregates of the eigenstates of the blankets of first level particles. Here, there were 1024 such eigenstates that have been partitioned into 57 particles. Each particle has one or more eigenstates; here, a total of 296. At this level, time constants have started to increase, including some particles that evince slow dynamics of about 10 seconds. Note, that the particles now have a greater spatial scale and have – in most instances – a symmetric spatial deployment across hemispheres. This reflects the fact that Jacobian includes transcallosal or interhemispheric coupling. For example, the first particle has one internal state (by design), 29 active states and 44 sensory states. These different states are colour-coded with white, light grey and dark grey, respectively – to illustrate the characteristic 'fried egg' arrangement in which internal states (white) are surrounded by blanket (i.e., active and sensory) states (in grey). The eigenmodes of this particle covers voxels in primary visual and extrastriate cortex. The second particle sits across the bilateral superior parietal cortices, while the third particle encompasses anterior (i.e., polar) temporal regions – and so on. The spatial scale of these particles corresponds roughly to a cytoarchitectonic parcellation. The ensuing (57) particles collectively comprise 296 eigenstates that are partitioned into five particles at the next level, corresponding roughly to lobar neuroanatomy.

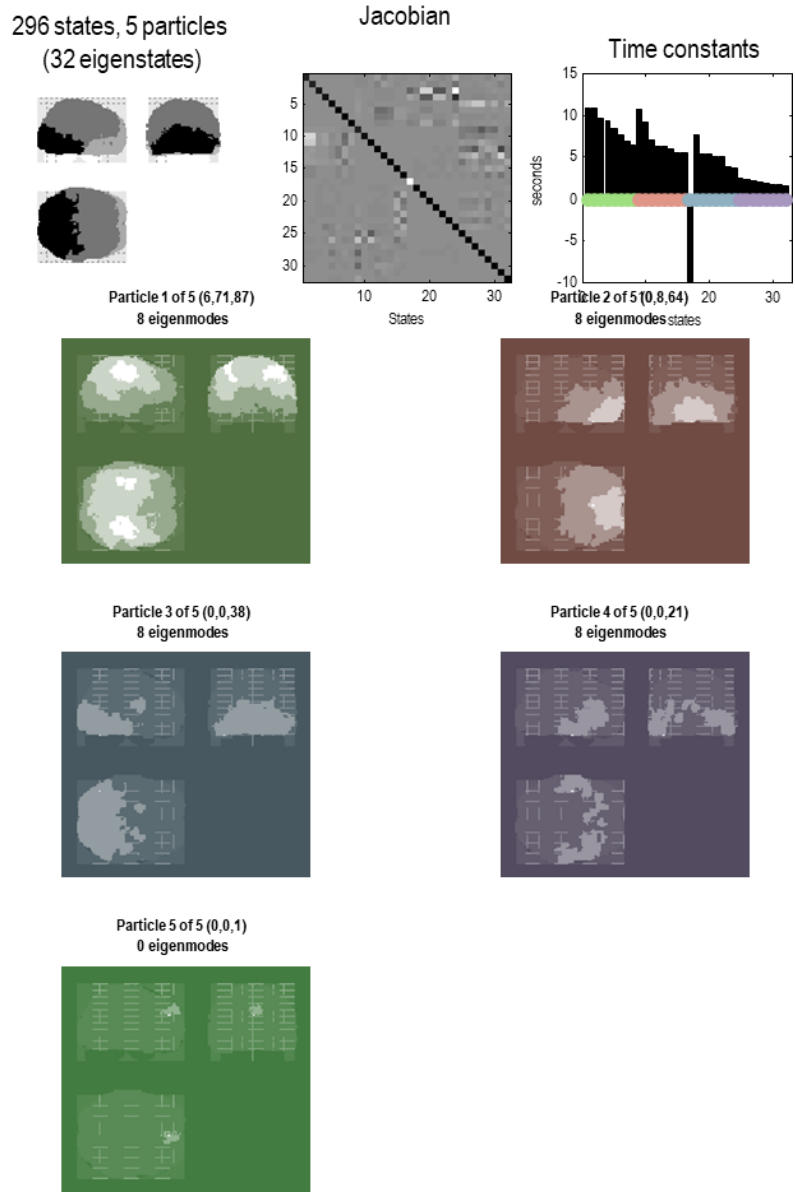

Technical note

**Particular partition at the third level**: this follows the same format as the previous figures. Here, the 57 particles from the previous scale are now partitioned into five particles that, collectively, possess 32 eigenstates. Here, the adjacency matrix is shown in image format, in terms of the real part of each (complex) Jacobian. At this scale, dynamics of each particle are becoming increasingly slow, with typical time constants between five and 10 seconds. The negative time constant reflects a positive eigenvalue that denotes an exponential *divergence* of trajectories that underwrites stochastic chaos. The five particles retain a degree of interhemispheric symmetry: the first particle has six internal states, 71 active states and 87 sensory states. This particle covers a large dorsal portion of cortex, including parietal cortex and frontal eye fields. Conversely, the second particle covers large regions of frontal cortex, with the eight active states located in orbitofrontal regions. The third particle is located in posterior visual and inferotemporal regions, while the fourth particle subsumes anterior temporal and ventral regions. Interestingly, there is one small particle (with a single sensory state) in the anterior medial prefrontal cortex. These five lobe-like particles (with 32 eigenstate's) now contribute to a single particle at the final (fourth) level.

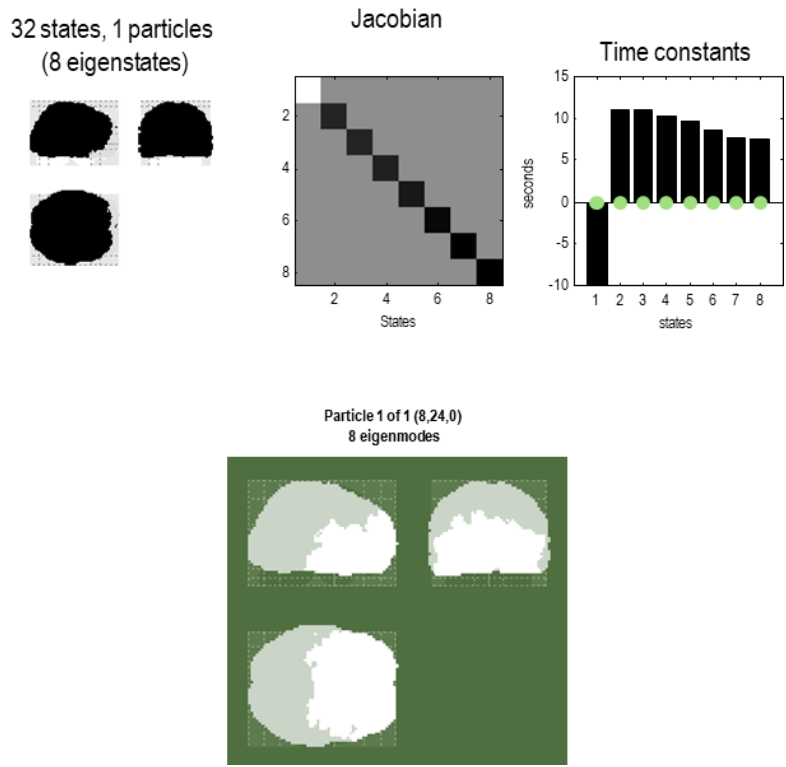

**Figure 8**

**Particular partition at the fourth scale**. The five particles of the previous level have now been partitioned into a single particle with eight internal states (in rostral regions) and 24 sensory states (in caudal regions), in white and light grey, respectively. This particle possesses eight eigenstates, the first of which has a positive eigenvalue – or negative time constant (denoting stochastic chaos). At this scale, all of the eigenstates have protracted dynamics, with time constants in the order of 10 seconds. Note that there is no coupling among the eigenstates in the Jacobian. This means the dynamics are completely characterised by the leading diagonal terms, i.e., the complex eigenvalues of the eight constituent eigenstates.

Note that it would have been possible to re-evaluate the Jacobian using another dynamic causal model of the eigenstates at any particular level and then use Bayesian model reduction to eliminate redundant coupling parameters. This is an interesting alternative to using the estimates of the Jacobian based upon the first-order approximation at the smallest scale. We will explore the impact of re-evaluating the Jacobian in subsequent work. For the purposes of the current illustration, we will retain the linear solutions at higher scales – based upon the lowest scale – to illustrate that one can still reproduce the emergent phenomena of interest described below. These dynamical phenomena are therefore directly attributable to local linear coupling with a particular sparsity structure that is sufficient to produce interesting self-organised dynamics at higher scales. Before taking a closer look at dynamics over scales, this section concludes with a brief characterisation of coupling at the smallest scale.

## Sparse coupling

The Jacobian from the above analysis summarises the effective connectivity at the smallest scale, where each node particle has a reasonably precise anatomical designation. This means that we can interpret the elements of the Jacobian in terms of directed (effective) connectivity. We had anticipated that this would mirror the exponential distance rule established through anatomical tracing studies (Finlay, 2016; Horvát et al., 2016; Wang & Kennedy, 2016). However, it did not. Instead, this (linear) characterisation of effective connectivity was better explained with a power law that, interestingly, was quantitatively distinct for inhibitory (i.e., negative) and excitatory (i.e., positive) connections (i.e., elements of the Jacobian).

Figure 9 summarises the statistical characteristics of coupling among particles at the first level. The upper left panel shows each connection in terms of the real part of the corresponding Jacobian in Hz, against the distance spanned by the connection (i.e., Euclidean distance between the centres of the two particles). Two things are evident from this scatterplot: first, positive (excitatory – red dots) connections dominate in the short range (around 8 mm), while negative (inhibitory – blue dots) dominate around 16 mm. Although there is variability, the dependency of the coupling strength on distance shows some lawful behaviour that is disclosed by plotting the log-coupling (real part) against log-distance (upper right panel). The circles are the averages in bins (discrete ranges) of the dots in the upper left panel. A linear regression suggests a scaling exponent of -1.14 for excitatory coupling and a smaller scaling exponent of -0.52, for inhibitory coupling. This dissociation is consistent with a Mexican Hat-like coupling kernel, with short range excitation and an inhibitory penumbra. This kind of architecture predominates in neural field models of cortical and subcortical coupling; e.g., (Coombes, 2005; Itskov, Curto, Pastalkova, & Buzsaki, 2011; Moran, Pinotsis, & Friston, 2013).

Technical note

The lower panel plots the strength of reciprocal connections against each other, to illustrate the relative proportions of recurrent excitatory and inhibitory coupling; here 65% and 31% respectively. There are only about 4% of connections that show an anti-symmetry, i.e., excitatory in one direction and inhibitory in the other. The rarefied region in the centre of this scatterplot reflects the fact that connections with small coupling strengths have been eliminated during Bayesian model reduction (see Figure 4). In the next section, we will see how this sparse local coupling engenders progressively more structured and itinerant dynamics at increasing spatial and temporal scales.

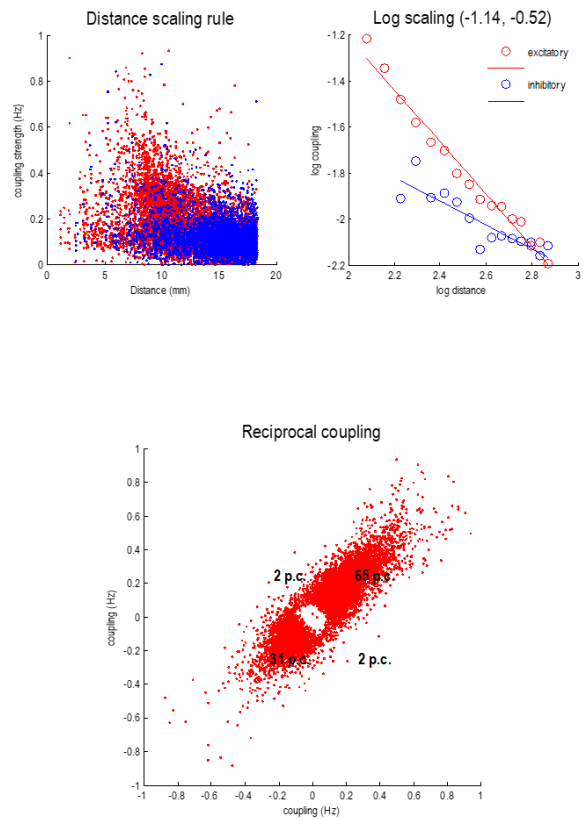

**Figure 9**

**Local connectivity**: This figure reports some of the statistics of dynamical coupling among particles at the first level. The **upper left panel** plots each connection in terms of the real part of the Jacobian in Hz, against the distance spanned by the connection. Two things are evident from this scatterplot: first, positive (excitatory – red dots) connections dominate in the short range (around 8 mm), while negative (inhibitory – blue dots) dominate around 16 mm. The **upper right panel** plots the log-coupling (real part) against log-distance, where circles report the averages in bins (discrete ranges) of the dots in the left panel. A linear regression gives a scaling exponent of -1.14 for excitatory coupling and a scaling exponent of the -0.52, for inhibitory coupling. The **lower panel** plots the strength of reciprocal connections against each other, to illustrate the relative proportions of recurrent excitatory and inhibitory coupling; here 65% and 31% respectively.

## Intrinsic dynamics

This section focuses on the intrinsic dynamics of each particle at different scales by associating the Jacobian of each particle with Lyapunov exponents. For people not familiar with dynamical systems theory, the Lyapunov exponents score the average exponential rate of divergence or convergence of trajectories in state space (Lyapunov & Fuller, 1992; Pavlos et al., 2012; Yuan, Ma, Yuan, & Ao, 2011). Because we are dealing with a linearised system, the Lyapunov exponents are the same as the eigenvalues of the Jacobian describing intrinsic coupling. By construction, this is a leading diagonal matrix containing intrinsic eigenvalues whose real values are close to zero. In terms of a linear stability analysis, the real part of these eigenvalues (i.e. self-induced decay) corresponds to the rate of decay. This means that as the eigenvalue approaches zero from below, the pattern of activity encoded by this eigenstate decays more and more slowly. This is the essence of critical slowing and means that, from the point of view of dynamical stability, this eigenstate has become unstable (Haken, 1983; Jirsa, Friedrich, Haken, & Kelso, 1994; Mantegna & Stanley, 1995; Pavlos et al., 2012). The complement of critical instability is a stable fast eigenstate that decays very quickly, i.e., an eigenstate whose eigenvalue has a large negative real part.

The imaginary part of the eigenvalue describes the characteristic frequency at which this decay is manifest. If the imaginary part is zero, the system decays monotonically. However, complex values mean that the intrinsic dynamics acquire a sinusoidal aspect. Because each particle has a number of eigenstates, an ensemble of particles can be construed as loosely coupled phase oscillators (M Breakspear & Stam, 2005; De Monte, d'Ovidio, & Mosekilde, 2003; Kaluza & Meyer-Ortmanns, 2010; Kayser & Ermentrout, 2010), featuring multiple frequencies. The associated dynamics of a single particle can be visualised by plotting its eigenvalues in the complex number plane. The closer the eigenvalue to the vertical axis, the slower the dynamics; such that as the real eigenvalue approaches zero (i.e., from the left), the particle approaches a transcritical bifurcation (at zero) and displays a simple form of critical slowing.

This characterisation of intrinsic dynamics – at different scales – is illustrated in the right panels of Figure 10. Note that the complex values are symmetrically paired (dots of the same colour). The key thing to observe here is that when we look at the eigenvalues of particles at higher scales, there are some eigenstates that approach criticality and start to show intrinsic oscillatory behaviour. This is one of the key observations from the current renormalisation scheme; namely, there is a necessary slowing as one progresses from one scale to the scale above. Furthermore, at higher scales intrinsic dynamics start to appear with progressively slower frequencies.

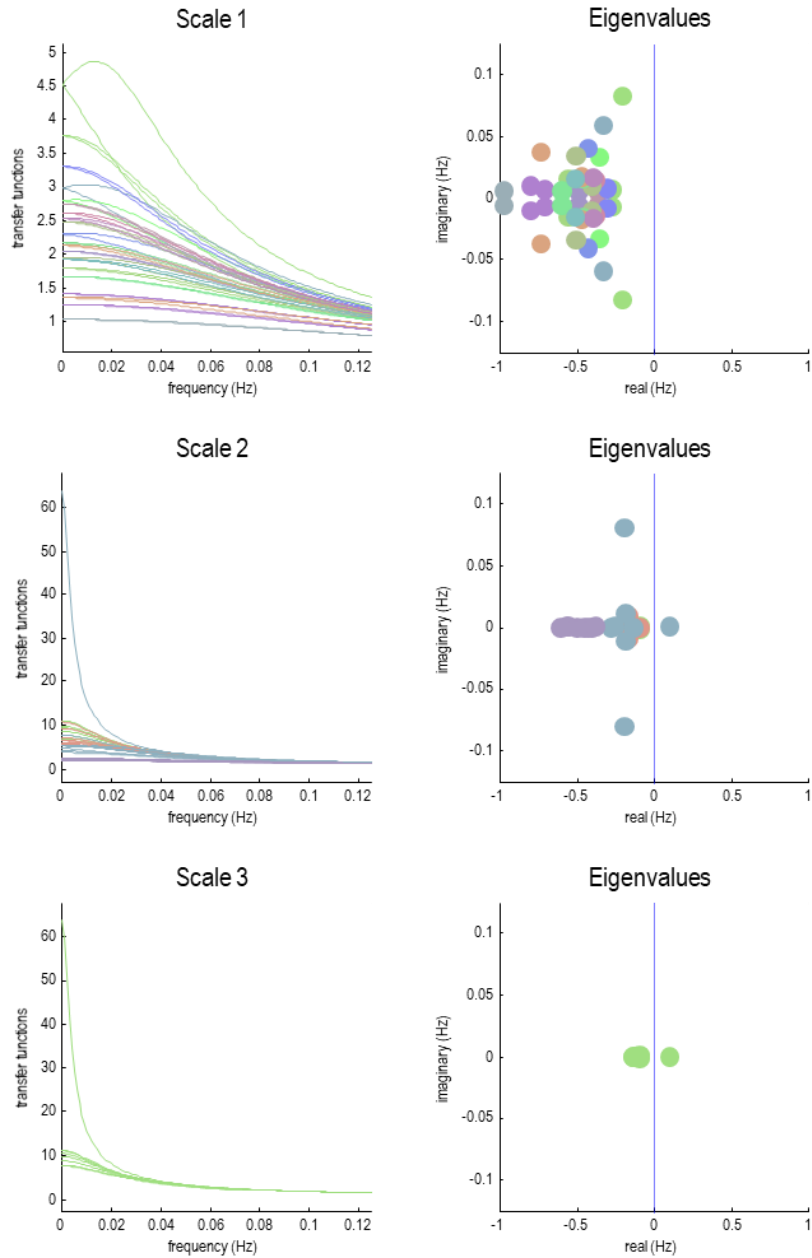

**Figure 10**

**Transfer functions**: This figure characterises the dynamics at successive scales in terms of transfer functions, as quantified by the complex eigenvalues (c.f., a pole-zero map). The **left column** shows the transfer functions of frequency for all particles with a complex eigenvalue (at successive scales). These eigenvalues are shown in the **right column** in the complex number plane. As we ascend from one scale to the next, the real part of the eigenvalue approaches zero from the left – and the number of eigenvalues (i.e., number of particles) falls with the coarse-graining. The complex part of an eigenvalue corresponds to the peak frequency of the associated transfer function, while the dispersion around this peak decrease as the real part approaches zero. The emergence of spectral peaks in the transfer functions inherit from the complex part of the eigenvalues – that emerge under asymmetric coupling with solenoidal flow. The next figure addresses the question: do these solenoidal dynamics vary in a systematic way over the brain?

Another way of characterising temporal dynamics is to use linear systems theory to map the eigenvalues to the spectral density of the timeseries that would be measured empirically. This rests upon standard transforms and the convolution theorem that enables us to express the systems first-order kernels as a function of the Jacobian (K. J. Friston, Kahan, et al., 2014). In frequency space, these kernels correspond to transfer functions and describe the spectral power that is transferred from the random fluctuations to the macroscopic dynamics of each eigenstate. The left panels of Figure 10 shows the transfer functions of the eigenstates of each particle at different scales. At the lowest scale, power is spread over a large range of frequencies. At progressively higher scales, the power becomes more concentrated in the lower frequencies with a transfer function that has a characteristic Lorentzian form. Crucially, the frequencies at the highest scale correspond to the characteristic ultra-slow frequencies studied in resting state fMRI; namely, $< 0.01$ Hz. This is an interesting observation, which suggests that one can explain ultra-slow fluctuations in resting state fMRI purely in terms of local directed coupling among small particles of brain tissue. Note that this explanation does not involve any haemodynamics because the Jacobian that gives rise to these slow oscillations pertains to the neuronal states (prior to haemodynamic convolution). In other words, this is not an artefact of removing fast frequencies from the measured fMRI signals.

One might ask if there is any systematic variation of these ultra-slow frequencies across the brain. Figure 11 reports the implicit intrinsic timescales at intermediate scales (second scale – upper rows: third scale – lower rows). The left column shows the eigenmodes in terms of their principal frequency, i.e., the largest complex eigenvalue (divided by $2\pi$). The right column shows the corresponding eigenmodes in terms of their principal time constants, i.e., the reciprocal of the largest negative real part. These two characterisations – principal frequency and time constant – speak to different aspects of intrinsic timescales; both of which contribute to the shape of an eigenstate's transfer function. The first quantifies the frequency of solenoidal flow, while the second reflects the rate of decay associated with the dissipative flow (we will unpack solenoidal and dissipative flows in the last section).

It is clear from these results that caudal (i.e., posterior) regions have faster intrinsic frequencies, relative to rostral (i.e., anterior) regions. Interestingly, in this example, the inferotemporal and ventral eigenmodes also show a relatively high frequency. At the third scale, this caudal-rostral gradient is more evident, suggesting that faster solenoidal dynamics dominate in posterior parts of the brain. This is consistent with both theoretical and empirical findings suggestive of a gradient of timescales – as one moves from the back to the front of the brain and, implicitly, from hierarchically lower areas to higher areas (Cocchi et al., 2016; Hasson, Yang, Vallines, Heeger, & Rubin, 2008; Kiebel, Daunizeau, & Friston, 2008; Liegeois et al., 2019; Murray et al., 2014; Wang & Kennedy, 2016). Note that the frequencies in question here are very slow; namely, about 0.01 Hz or below. These are the ultra-slow frequencies typically characterised in resting state fMRI (Liegeois et al., 2019). In the present setting, these ultra-slow frequencies are an emergent property of scale-invariant behaviour, when one moves from spatial temporal scales suitable for describing lobar dynamics or large intrinsic brain networks.

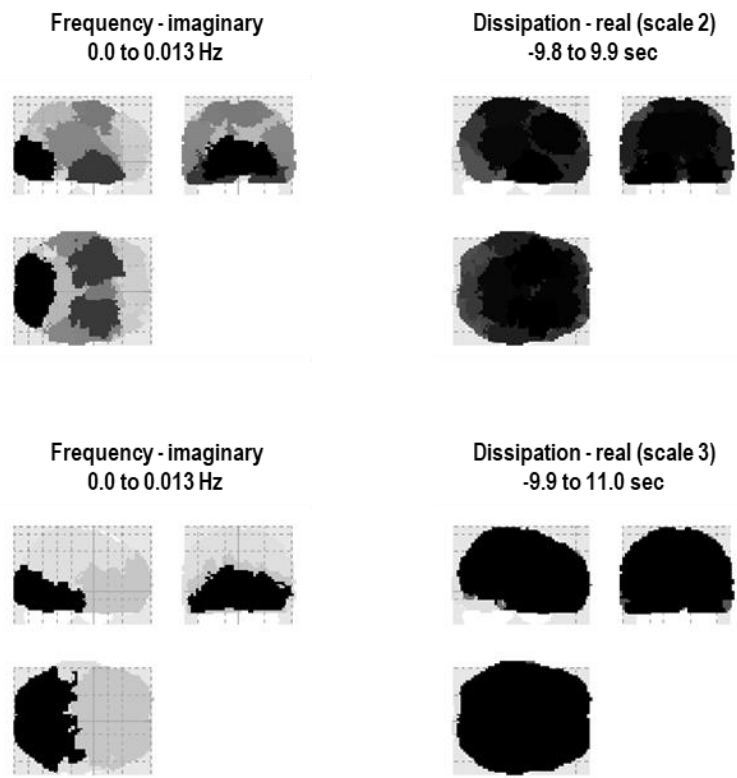

**Figure 11**

**Intrinsic timescales in the brain**: This figure reports intrinsic timescales at intermediate scales (second scale – upper rows: third scale – lower rows). The **left column** shows the eigenmodes in terms of their principal frequency, i.e., the largest complex eigenvalue (divided by $2\pi$). The **right column** shows the corresponding eigenmodes in terms of their principal time constants, i.e., the reciprocal of the largest negative real part.

The eigenvalues in Figure 10 take positive real values at higher (second and third) scales. This means that they have crossed the zero threshold to engender a transcritical bifurcation. Strictly speaking, these produce solutions that cannot be realised because of an exponential divergence of trajectories. This reflects the first-order approximation that we are using to summarise the dynamics. Although this linear approximation precludes stochastic chaos, positive real values speak to the notion that some particles at higher scales become excitable for short periods of time. This means that we are moving away from a loosely coupled oscillator model – that has a fixed point or limit cycle attractor – towards what physicists call active or excitable matter (Keber et al., 2014; Ramaswamy, 2010). This is a nice metaphor for the brain and means that if the particles that constitute active (grey) matter are considered in isolation, they can show autonomous dynamics that can be characterised as stochastic chaos or itinerancy. Indeed, one can use the Kaplan-Yorke conjecture to associate these itinerant excitable autonomous dynamics with a correlation dimension using the following equality (Kaplan & Yorke, 1979):

$$D^{(i)} = k + \sum_{j=1}^{k} \frac{\lambda_j^{(i)}}{|\lambda_{k+1}^{(i)}|} \tag{10}$$

Here, $\lambda_1^{(i)} \geq \cdots \geq \lambda_k^{(i)}$ are the $k$ largest real exponents (i.e., intrinsic eigenvalues), for which the sum is non-negative. One can see here that certain particles would attain a fractional (non-integer) correlation dimension based on this conjecture, i.e., a fractal aspect.

At this point we can return to the renormalisation group and **RG** scaling behaviour. This scaling behaviour depends upon the link between various parameters of the systems Lagrangian (or equivalent characterisation of dynamics) between successively higher levels. Consider the following **RG** flow or beta function as an instance of equation (4):

$$\sigma_\tau^{(i+1)} = e^{\beta_\tau} \sigma_\tau^{(i)}$$
$$\sigma_\tau^{(i)} = E[-\text{Re}(\lambda_{nn_j}^{(i)})]$$

This says as we move from one scale to the next, the time scale increases by $e^{\beta_\tau} \geq 1$. Invoking the same beta function for spatial scale induces a relationship between temporal and spatial scales in the form of a power law with a scaling exponent $\alpha$. This exponent corresponds to the ratio of the time and spatial exponents of the beta functions:

$$\left.\begin{array}{l}\sigma_\tau^{(i+1)}=e^{\beta_\tau}\sigma_\tau^{(i)}\\\sigma_\ell^{(i+1)}=e^{\beta_\ell}\sigma_\ell^{(i)}\end{array}\right\}\Rightarrow\left\{\begin{array}{l}\sigma_\tau^{(i)}=e^{i\beta_\tau}\sigma_\tau^{(0)}\\\sigma_\ell^{(i)}=e^{i\beta_\ell}\sigma_\ell^{(0)}\end{array}\right\}\Rightarrow\ln\frac{\sigma_\tau^{(i)}}{\sigma_\tau^{(0)}}=\alpha\ln\frac{\sigma_\ell^{(i)}}{\sigma_\ell^{(0)}}\Rightarrow$$

$$\sigma_\tau^{(i)}=\gamma(\sigma_\ell^{(i)})^\alpha$$

$$\alpha=\frac{\beta_\tau}{\beta_\ell},\ \gamma=\sigma_\tau^{(0)}(\sigma_\ell^{(0)})^{-\alpha}$$

(11)

The last quality in the first line follows by eliminating the scale *i* from the pair of beta functions. Intuitively, this scaling behaviour means that as we move from one scale to the next, things get slower and bigger – but at different geometric rates. This difference gives rise to a scaling exponent that links the increases in spatial scale to increases in temporal scale. We evaluated the characteristic time and spatial constants for each scale by taking the mean of the real eigenvalues and the spatial dispersion of the corresponding eigenmodes associated with all particles at each scale:

$$\sigma_\tau^{(i)}=E[-\operatorname{Re}(\lambda_{nn_j}^{(i)})]$$

$$\sigma_\ell^{(i)}=E[|\,v_{n_j}^{(i)}\,|^{1/3}]$$

(12)

Plotting the logarithms of these values against each other allows one to estimate the scaling exponent using linear regression. Figure 12 shows the results of this analysis across all scales. The scaling exponent here was 1.14. This is not dissimilar to the value of 1.47 obtained with a similar analysis of murine calcium imaging data (Fagerholm et al., 2019), where coarse graining was implemented by averaging over spatial blocks. To put this value into perspective, the scaling exponent for Kepler's laws of motion is 1.5. This scaling exponent reflects the disparity in spatial and temporal constants; where the temporal constant increases by a factor of 2.37 from one scale the next, while the spatial support increases by 2.13.

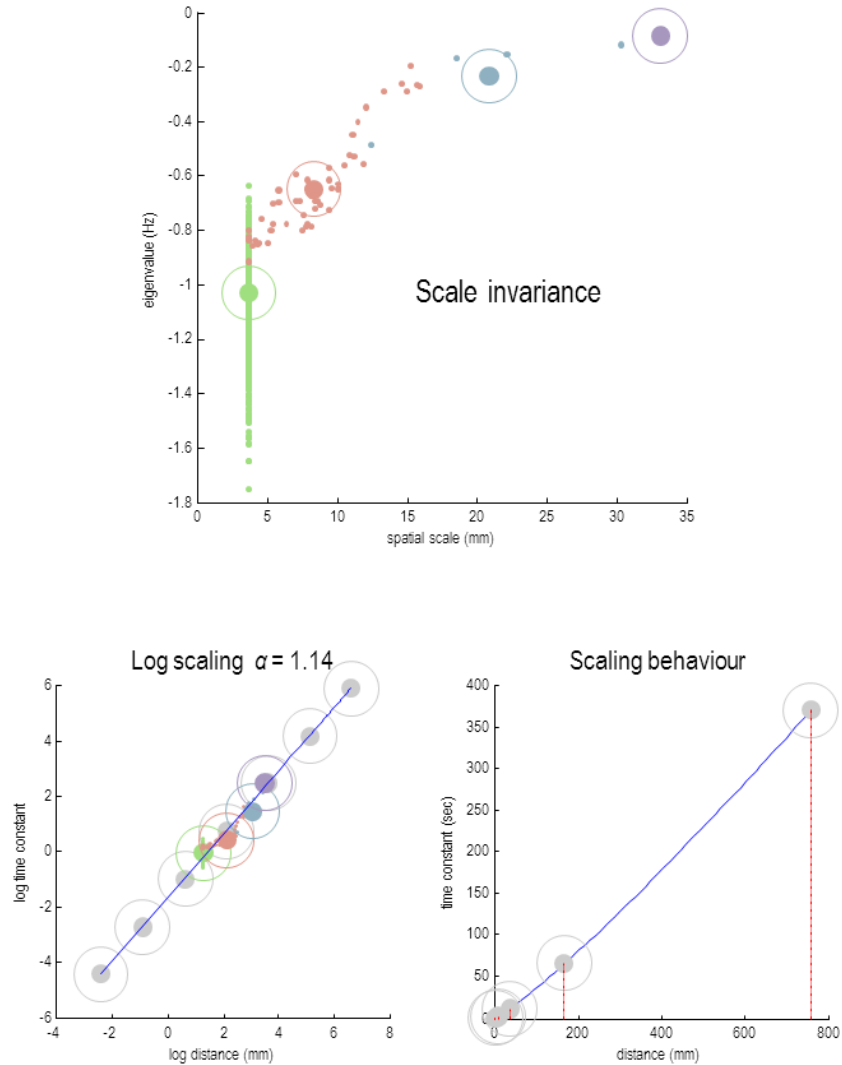

**Scale invariance**: This figure illustrates scaling behaviour across the scales of the particular decomposition. The **upper panel** plots the real part of the eigenvalues of each particle against its spatial scale; namely the calliper width of each particle's eigenmode. This is replicated for each of the four scales, denoted by the different colours (green, pink, cyan and puce, respectively). The expected values are shown as encircled large dots. The lower left panel plots the logarithms of these temporal and spatial expectations against each other. The resulting regression slope corresponds to the scaling exponent; here, 1.14. The light grey circles correspond to what would have been seen at higher and lower scales. The **lower right plot** shows the same regression in terms of the implicit time constant, as a function of spatial scale expressed in millimetres. The red lines correspond to the largest scales (of $i = 6$ and $i = 8$) depicted in the left panel – suggesting characteristic time constants in the order of 60 and 360 seconds. This scaling behaviour suggests that as we increase the spatial scale or coarse graining, dynamics become slower, as the real parts of particular eigenvalues approach zero from below.

Technical note

This scale free behaviour means that we can evaluate the time constants at scales that we have not characterised empirically. Table 1 lists these extrapolated or projected timescales right down to the nanoscale and up to higher scales that would be appropriate to talk about networks of brains or social communities or institutions.

<div align="center">

**Table 1**

</div>

<div align="center">

Spatiotemporal scales and examples

</div>

| Scale | Spatial scale | Time scale | Example |
|-------|---------------|------------|---------|
| -8 | 4.38 μm | 380 μs | Dendritic spines occur at a density of up to 5 spines per μm of dendrite. Spines contain fast voltage-gated ion channels with time constants in the order of 1 ms. |
| -4 | 89.3 μm | 11.9 ms | A cortical minicolumn: a minicolumn measures of the order of 40–50 μm in transverse diameter 80 μm spacing (Peters & Yilmaz, 1993). The membrane time constant of a typical cat layer III pyramidal cell is about 20 ms. |
| 0 | 1.82 mm | 374 ms | A cortical hypercolumn (e.g., a 1 mm expense of V1 containing ocular dominance and orientation columns for a particular region in visual space (Mountcastle, 1997)). Typical duration of evoked responses in the order of 1 to 300 ms (c.f., the cognitive moment). |
| 4 | 37.2 mm | 11.8 sec | The cerebellum is about 50 mm in diameter, corresponding to the size of cortical lobes. Sympathetic unit activity associated with Mayer waves within frequency of 0.1 Hz (wavelength of 10 seconds). |
| 8 | 758 mm | 6.15 min | A dyadic interaction (e.g., a visit to your doctor). |
| 12 | 15.5 m | 3.22 hrs | A dinner party for six guests, lasting for several hours. |
| 16 | .31 km | 4.21 days | An international scientific conference (pre-coronavirus). |

This completes our discussion of scale invariance and associated dynamics, where we have taken a special interest in the temporal scaling behaviour that emerges from local connectivity at smaller scales of analysis. In the next section, we turn to the coupling between particles and see what this has to say in terms of how intrinsic brain networks influence each other.

## Extrinsic dynamics

In this section, we consider the off-diagonal elements of the Jacobian at the successive scales afforded by the renormalisation group. By construction, these terms couple different particles. The *ij*-th element of the *nm*-th block of the Jacobian couples the *j*-th eigenstate of the *m*-th particle to the *i*-th eigenstate of the *n*-th particle.

$$J = \begin{bmatrix} \lambda_{11}^{(i)} & \cdots & \lambda_{1n}^{(i)} \\ \vdots & \ddots & \vdots \\ \lambda_{n1}^{(i)} & \cdots & \lambda_{nn}^{(i)} \end{bmatrix} = \begin{bmatrix} \lambda_{11_{11}}^{(i)} & \cdots & 0 & \lambda_{1n_{11}}^{(i)} & \cdots & \lambda_{1n_{1j}}^{(i)} \\ \vdots & \ddots & \vdots & \cdots & \vdots & \ddots & \vdots \\ 0 & \cdots & \lambda_{11_{jj}}^{(i)} & \lambda_{1n_{j1}}^{(i)} & \cdots & \lambda_{1n_{jj}}^{(i)} \\ \vdots & & \vdots & & \ddots & & \vdots \\ \lambda_{n1_{11}}^{(i)} & \cdots & \lambda_{n1_{1j}}^{(i)} & \lambda_{nn_{11}}^{(i)} & \cdots & 0 \\ \vdots & \ddots & \vdots & \cdots & \vdots & \ddots & \vdots \\ \lambda_{n1_{j1}}^{(i)} & \cdots & \lambda_{n1_{jj}}^{(i)} & 0 & \cdots & \lambda_{nn_{jj}}^{(i)} \end{bmatrix}$$

(13)

This directed coupling is generally complex. The complex part can be thought of as inducing a phase shift or delay in the influence of one eigenstate on another. The real part is of more interest here and corresponds to a rate constant; much like the real part of the Lyapunov exponents of the intrinsic coupling describe the rate of decay. However, here, we are talking about the rate at which an eigenstate of one particle responds to the eigenstate of another. This means that large positive or negative real extrinsic coupling become interesting (previously, we have been discarding eigenstates with large negative intrinsic eigenvalues because they dissipate almost immediately). Figure 13 illustrates this extrinsic (between particle) coupling at the penultimate scale (scale three) in the form of a connectogram.

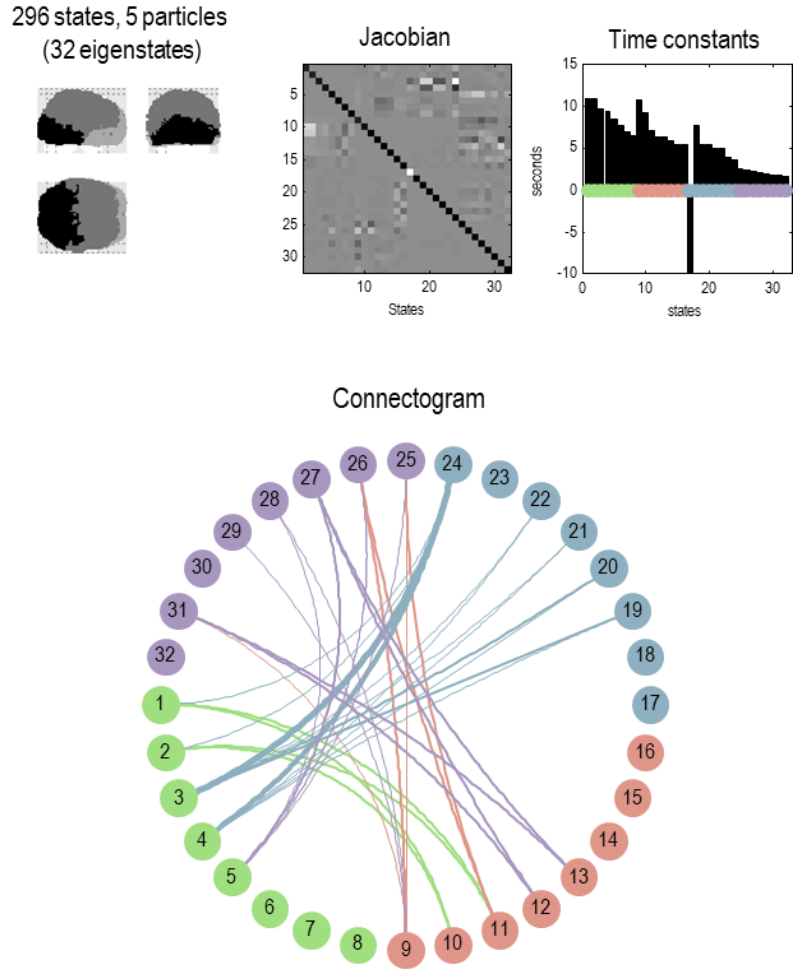

**Figure 13**

**Extrinsic connectivity**: This figure illustrates asymmetric extrinsic (between particle) coupling at the penultimate scale (scale three). The **upper panels** reproduce the results in Figure 7, while the **lower panel** is a connectogram illustrating the coupling among the (32) eigenstates that constitute the five particles at this scale. The width of each connector reflects the strength of the coupling – after dividing the strength into five bins and eliminating the lowest bin. The colour of the dots corresponds to the colour of the particle in the upper right panel. The colour of the connectors corresponds to the source of the strongest (reciprocal) connection. In this example, the largest afferent connection is from eigenstate 24 to eigenstate three. This corresponds to an influence of the first eigenstate of the fourth (cyan) particle on the third eigenstate of the first (green) particle. The coupling strength corresponds to the real part of the Jacobian, in Hz. The fact that coupling is mediated by complex coupling coefficients means that the influence of one eigenstate on another can show profound asymmetries in time. This is illustrated in the next figure – that examines the largest connection above in more detail.

The implications of complex extrinsic coupling can be understood in terms of cross-covariance functions of time that characterise delayed or lagged dependencies[5]. For example, Figure 14 characterises these dependencies between the two eigenstates with the strongest coupling at this (third) scale. The implicit coupling is mediated by the corresponding element of the (complex) Jacobian – circled in red in the upper middle panel. The flanking panels on the left and right show the associated eigenmodes in voxel space. The middle row shows the auto-covariance functions of the two eigenstates, illustrating serial correlations that can last for many seconds. The interesting part of this figure is in the lower panels: these report the cross-covariance function between the two eigenstates, over 256 seconds (lower left panel) and 32 seconds (lower right panel), respectively. The key thing to observe here is that the peak cross-covariance emerges at an eight second lag from the 24th to the third eigenstate. This asymmetrical cross-covariance (and implicitly cross-correlation) function reflects the solenoidal coupling and implicit breaking of detailed balance accommodated by the particular decomposition (see next section). Note that the (zero lag) correlation is almost zero. This speaks to the potential importance of using cross-covariance functions (or complex cross spectral in frequency space), when characterising functional connectivity in distributed brain responses (K. J. Friston, Bastos, et al., 2014; Mohanty, Sethares, Nair, & Prabhakaran, 2020). This brief treatment of extrinsic coupling has made much of the complex nature of dynamical coupling and how it manifests in terms of functional connectivity. In the final section, we revisit this kind of coupling in terms of nonequilibrium steady states.

---

[5] The cross-covariance functions can be evaluated in a straightforward from the complex transfer functions, shown in Figure 10. In other words, they can be derived directly from the Jacobian, under first-order assumptions.

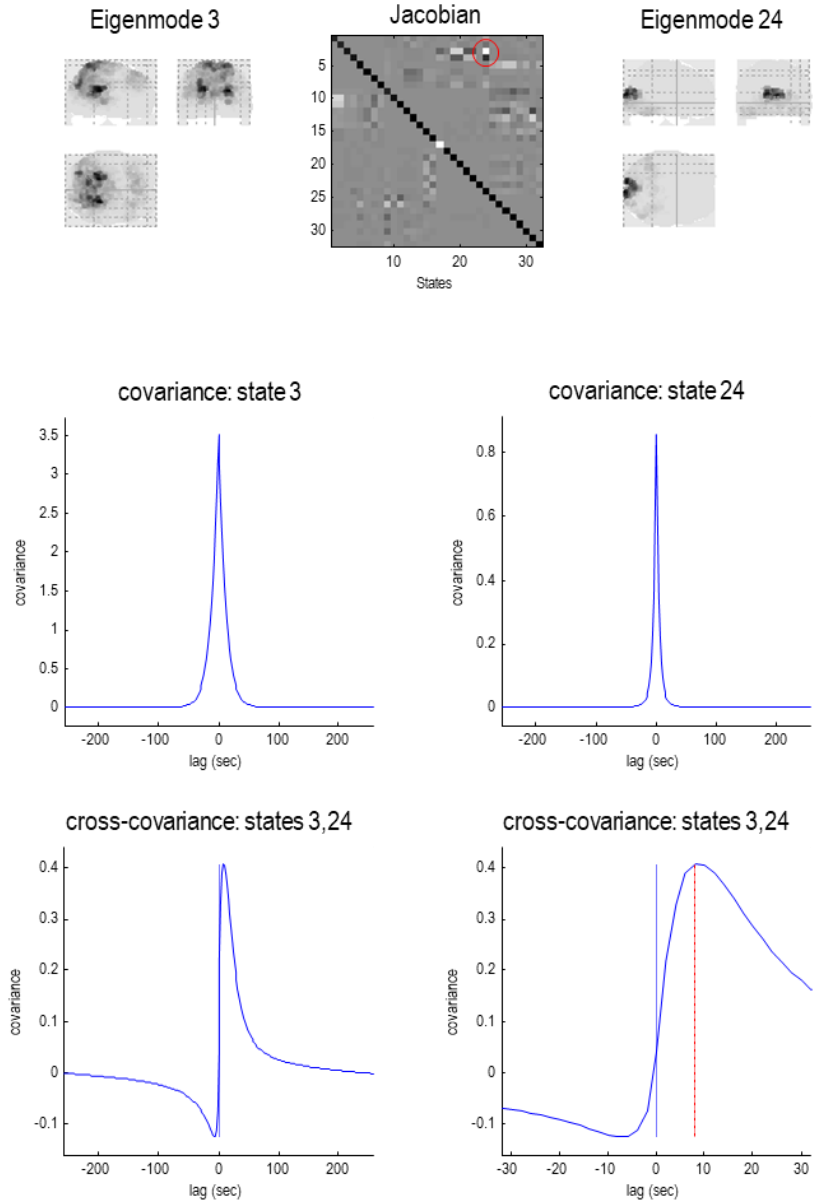

**Figure 14**

**Dynamic coupling**: This figure characterises the coupling between the two eigenstates of the previous figure with the strongest coupling at the third scale. This coupling is mediated by the corresponding element of the (complex) Jacobian – circled in red in the **upper middle panel**. The flanking panels on the **left** and **right** show the corresponding eigenmodes in voxel space. The **middle row** shows the auto-covariance functions of the two eigenstates, illustrating serial correlations that can last for many seconds. The **lower two panels** report the cross-covariance function between the two eigenstates, over 256 seconds (**lower left panel**) and 32 seconds (**lower right panel**). The red line indicates the peak cross covariance at about eight seconds lag.

# Dynamics and statistical physics

Above, we have referred to solenoidal and dissipative flows, in relation to the complex and real parts of intrinsic eigenvalues – and how they manifest in terms of intrinsic brain networks. This section unpacks this relationship by applying the statistical physics of nonequilibrium steady states to neuronal fluctuations. Our focus will be on the relationship between conventional characterisations of functional connectivity and the more general formulation afforded by a particular decomposition. Specifically, we will see that conventional formulations assume a special case that discounts solenoidal flow – and implicitly assumes neuronal dynamics attain steady state at statistical equilibrium.

In the previous section, we examined extrinsic coupling among particles in terms of their covariance. Here, we return to coupling and dynamics that are intrinsic to a particle; namely, the final particle at the last level. In this example, the particle has eight eigenstates, whose complex eigenvalues imply a loss of detailed balance and implicit steady state that is far from equilibrium. To understand the link between detailed balance and equilibria versus nonequilibrium steady states, it is useful to consider the eigen-decomposition of the final particle in relation to standard analyses of functional connectivity (e.g., singular value decomposition or principal component analysis of covariance matrices). In what follows, we first rehearse the relationship between flow and steady state distributions over states afforded by the Helmholtz decomposition. We then look at what this implies under the assumption of symmetric coupling – and how this leads to equilibrium mechanics and a simple relationship between the Jacobian and covariances among their respective eigenstates. We then revisit these relationships but replacing solenoidal flow, to clarify the differences between summarising dynamics in terms of the eigenvectors of the Jacobian and the eigenvectors of the functional connectivity matrix.

In general, one can express the flow at steady state in terms of a Helmholtz decomposition of the solution to density dynamics (as described by the Fokker Planck equation). This is an important expression that underwrites much of physics and related treatments of self-organisation in the biological sciences[6]. Starting with a Langevin formulation of neuronal dynamics, we can express the flow of states in equation (2) as

---

follows[7]:

$$\dot{x} = f(x) + \omega$$
$$f(x) = (Q - \Gamma)\nabla \Im(x)$$

$$(14)$$

Here, $\Im(x) = -\ln p(x)$ is a potential energy that quantifies the surprise at finding the brain in any state. The positive definite matrix $\Gamma \propto I$ plays the role of a diffusion tensor describing the amplitude of random fluctuations, $\omega$ (assumed to be a Wiener process), while the antisymmetric matrix $Q = -Q^{\dagger}$ mediates solenoidal flow. Equation (14) says that the expected flow at any point in state space has two components: a dissipative gradient flow, $-\Gamma \nabla \Im$ on the logarithm of the steady state density and a solenoidal flow, $Q \nabla \Im$ that circulates on the isocontours of this density. In brief, the gradient flow counters the dispersive effects of random fluctuations, thereby rendering the probability density stationary. On differentiating the Helmholtz decomposition, with respect to systemic states we have, $\forall x$:

$$f(x) = (Q - \Gamma)\nabla \Im(x) \Rightarrow J(x) = (Q - \Gamma)\Pi(x)$$

$$(15)$$

Here, the Jacobian $J = \nabla f(x)$ and Hessian $\Pi(x) = \nabla^2 \Im(x)$ are functions of states. Because the Hessian matrix is symmetrical, there are linear constraints on the solenoidal coupling (Qian & Beard, 2005):

$$(Q - \Gamma)^{-1}J(x) = \Pi(x) = \Pi(x)^T = J(x)^T(Q - \Gamma)^{-T}$$
$$\Rightarrow$$
$$QJ(x)^T + J(x)Q = \Gamma J(x)^T - J(x)\Gamma$$

These constraints mean that in the absence of solenoidal coupling – when random fluctuations have the same amplitude everywhere – the Jacobian has to be symmetric $Q = 0 \Rightarrow \Gamma J(x)^T = J(x)\Gamma$. In other words, symmetric coupling guarantees detailed balance (i.e., an absence of solenoidal flow).

## Detailed balance and Heisenberg's uncertainty principle

So how does this help us connect conventional analyses of functional connectivity to the eigenvectors of the Jacobian? First, if we make the simplifying assumption that effective connectivity is symmetric, we can ignore solenoidal flow. If we make the further assumption that the steady state is Gaussian[8] the Hessian can be interpreted as a precision matrix (i.e., inverse covariance or functional connectivity matrix). Under these simplifying assumptions, the Jacobian becomes a scaled version of the precision: setting $Q = 0$ in equation (15) gives:

$$J(x) = -\Gamma \Pi(x)$$

$$(16)$$

---

[7] Here, we have omitted (correction) terms that generalise this (Helmholtz) decomposition, because we are assuming that the amplitude of random fluctuations and the solenoidal terms change slowly over state space. We have also dropped the scale superscripts for clarity.

This means that the eigenvalues of the Jacobian, which reflect the rate of dissipation of each mode, are inversely related to the eigenvalues of the precision matrix. In other words, if we were to perform a principal component analysis of the covariance matrix $\Sigma = \Pi^{-1}$, the principal eigenvalues would be interpreted as explaining the most variance in the eigenstates. However, this is exactly the same as identifying the eigenstates whose flow has the smallest rate constant. In other words, the principal components are just the slow, unstable modes that do not dissipate quickly.

$$\xi^- J \xi = \lambda$$
$$\xi^- \Sigma \xi = -\Gamma \lambda^{-1}$$

$$(17)$$

One can quantify the dissipative aspect of the eigenmodes in terms of the expected dispersion of the flow:

$$E[f \times f] = J \Sigma J^T = \Gamma \Pi \Gamma \Rightarrow E[f \times f]E[x \times x] = \Gamma^2$$
$$\Sigma = E[x \times x] = \Pi^{-1}$$
$$f = Jx$$

$$(18)$$

This expression shows that the uncertainty about the flow – over state space at steady state – is inversely proportional to the corresponding uncertainty about the state (i.e., variance). This is Heisenberg's uncertainty principle. The connection to the uncertainty principle can be made explicit by associating the amplitude of random fluctuations with inverse mass (Karl Friston, 2019), where the constant of

---

[8] This Gaussian assumption is usually motivated in terms of a first order approximation to the flow (in terms of the Jacobian) around the maxima of the steady-state density. To the extent that the steady state density approximates a Gaussian, then this local linear approximation becomes global.

proportionality is Planck's constant. Equation (18) can then be expressed as:

$$E[mf \times mf\,]E[x \times x] = \left(\tfrac{\hbar}{2}\right)^2$$

$$
\begin{aligned}
2\Gamma &= \tfrac{\hbar}{m} \\
\Im(x) &= \tfrac{1}{2}x \cdot \Pi \cdot x \\
f(x) &= -\tfrac{\hbar}{2m}\nabla\Im = -\tfrac{\hbar}{2m}\nabla^2\Im \cdot x \\
\Pi &= \nabla^2\Im
\end{aligned}
\tag{19}
$$

This can be interpreted as follows: if we are fairly certain about the state of a system, we will be very unsure about its flow – and *vice versa*. This follows from the fact that, at steady-state, systems with predictable, slow flows become dispersed over state space, in virtue of the random fluctuations. Conversely, if a system can "gather it states up" and locate them in a small regime of state space, the requisite flows must be fast and varied.

In summary, if we assume detailed balance (i.e., discount solenoidal flow), we are assuming an equilibrium steady state of the sort studied in quantum and statistical mechanics. In this special case, there is a direct relationship between the (eigenvectors of) the Jacobian and the Hessian matrix (i.e., precision, or inverse functional connectivity) matrix. Furthermore, there is also a direct relationship between the Jacobian and the variance of the expected flow or dynamics. The assumption of detailed balance is licensed in many situations. Particularly, if we are dealing with ensembles of states or particles that are *exchangeable* (e.g., an idealised gas). This renders the Jacobian symmetrical and ensures detailed balance. The Jacobian is symmetrical because the influence of one particle on a second, is the same as the influence of the second on the first. However, this symmetry cannot be assumed in biological systems that break detailed balance, especially the brain. We now rehearse the above analysis by retaining the symmetry breaking, solenoidal flow that underwrites non-equilibrium steady-state dynamics.

## Nonequilibrium steady states and solenoidal flow

In the presence of solenoidal flow, the eigenvectors of the Jacobian and Hessian are no longer the same. So, which is the best summary of dynamics? Clearly, there is no definitive answer to this question; however, if we are interested in relevant quantities 'that matter', we are specifically interested in non-dissipative, slow, unstable dynamics. By construction, this is what the particular decomposition 'picks out' – by discarding fast fluctuations at each successive scale. This means that the eigenstates of the final particle should have identified slow, unstable, or critical dynamics. In contrast, had we just taken the principal components of the covariance matrix of the data (i.e., functional connectivity), we may not have identified the slow modes.

This begs the question, to what extent do solenoidal dynamics contribute to the intrinsic dynamics of the final particle? One can evaluate the relative contribution of dissipative gradient flows and non-dissipative solenoidal flow in terms of their expected dispersion:

$$
\begin{aligned}
E[f \times f] &= J \Sigma J^T = (Q - \Gamma) \Pi (Q - \Gamma)^T \\
&= \Gamma \Pi \Gamma + \underbrace{Q \Pi Q^T - \Gamma \Pi Q^T - Q \Pi \Gamma}_{\text{non-dissipative}}
\end{aligned}
\tag{20}
$$

Clearly, to do this, we need estimates of the amplitude of intrinsic fluctuations and the solenoidal term. However, under local linear (i.e., Gaussian) assumptions, these two quantities must satisfy $JQ + QJ^T = J^T \Gamma - \Gamma J$ or in terms of eigenstates, $\lambda Q + Q \lambda^\dagger = \lambda^\dagger \Gamma - \Gamma \lambda$. We can use this constraint to decompose the *kinetic energy* of the flow in terms of, and only of, the eigenvalues of the Jacobian, where $\kappa = -Re(\lambda)$:

$$
\frac{m}{\hbar} E[f \times f^\dagger] = \underbrace{\frac{\overbrace{Re(\lambda)^2}^{\text{dissipative}} + \overbrace{Im(\lambda)^2}^{\text{non-disspative}}}{-2 Re(\lambda)}}_{\text{kinetic energy}} = \frac{\lambda^\dagger \lambda}{2 \kappa}
$$

$$
\begin{aligned}
\lambda &= \xi^- J \xi \\
\Gamma &= \xi^- \Gamma \xi = \frac{\hbar}{2m} \\
Q &= -i \frac{Im(\lambda)}{Re(\lambda)} \Gamma \Rightarrow \lambda Q + Q \lambda^\dagger = \lambda^\dagger \Gamma - \Gamma \lambda \\
\Pi &= -\frac{1}{\Gamma} Re(\lambda) \Rightarrow \lambda = (Q - \Gamma) \Pi
\end{aligned}
\tag{21}
$$

The first equality follows from substituting the subsequent equalities in equation (20). The use of kinetic energy here appeals to equation (19), in which the amplitude of random fluctuations is associated with inverse mass. This equality says that the dissipative part of flow is determined by the real part of an eigenstate's eigenvalue, while the solenoidal contribution is the imaginary part squared, divided by the real part. Intuitively, this would be like decomposing the kinetic energy of the Earth into a solenoidal component corresponding to its orbital velocity – and a dissipative component, as it is drawn towards the sun. This speaks to an increase in kinetic energy with the frequency of (e.g., neuronal) oscillations, which is not unrelated to the Plank-Einstein and de Broglie relations in physics.

Note that when working with eigenstates, the solenoidal terms are encoded by $Q$, which is a leading diagonal matrix of imaginary values. Similarly, the dissipative terms are encoded by $\Gamma$, which is a leading diagonal matrix of real values. In other words, nonequilibrium steady-state – as defined by the prevalence of solenoidal flow – manifests as the imaginary parts of the Helmholtz decomposition, when the system is projected onto the eigenvectors of the Jacobian.

Figure 15 shows the dissipative and solenoidal (kinetic) energy of the eigenstates at the final scale. The corresponding eigenmodes are shown in the subsequent panels as maximum intensity projections (of their absolute values). In terms of dissipative dynamics, the first eigenmode has the smallest dissipative energy. In other words, it features the slowest, most unstable mode macroscopic (intrinsic) dynamics. Eigenmodes two and three are a conjugate pair, with complex parts – and, implicitly, a solenoidal contribution to their (kinetic) energy. These nodes are most pronounced in dorsal mid-prefrontal regions. Note that the kinetic energy of the first eigenstate is negative. This may seem counterintuitive; however, it is a simple reflection of the fact that the principal eigenvalue has a real part that is greater than zero. Clearly, the implicit exponential divergence of trajectories cannot persist globally. In a more complete analysis, the (stochastic) trajectory would quickly enter regimes of dissipation, such that the average real part (c.f., Lyapunov exponent) was less than zero. One might ask where does the dissipative energy come from? It is effectively driven by intrinsic fluctuations that, at the lowest level include the fluctuations in active states, which play the role of experimental or sensory inputs. This raises an interesting question: at what scale do experimental inputs manifest?

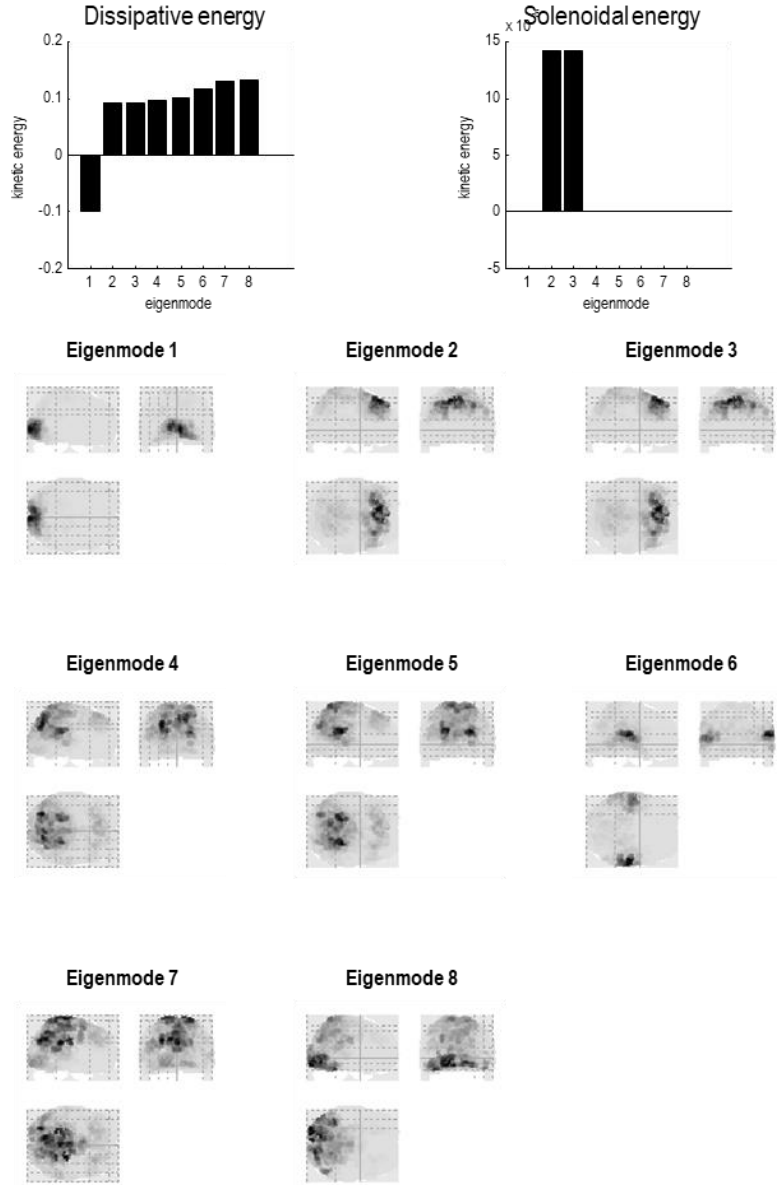

Figure 15

**Dissipative and solenoidal dynamics**: This figure unpacks the intrinsic coupling at the final (fourth, single particle) level. At this level, there can be no coupling between particles and – by construction – the dynamics is completely characterised in terms of the eigenstates that comprise the particle. In turn, these are completely characterised by their complex eigenvalues; namely, the intrinsic complex coupling. The upper panels show the dissipative and solenoidal (kinetic) energy of the eight eigenstates that comprise the particle. The corresponding eigenmodes are shown in the subsequent panels as maximum intensity projections (of their absolute values). In terms of dissipative dynamics, the first eigenmode has the smallest dissipative energy. In other words, it is the slowest, most unstable mode of this particle. Eigenmodes two and three are a conjugate pair, with complex parts – and, implicitly, a solenoidal contribution to their (kinetic) energy. These nodes are most pronounced in dorsal mid-prefrontal regions, with some expression in posterior parietal regions. The dissipative energy is, effectively, driven by intrinsic fluctuations that, at the lowest level include the fluctuations in active states, which play the role of experimental or sensory inputs.

## Dissipative brain responses

An intuitive way of thinking about the distinction between dissipative and solenoidal dynamics is in terms of the fluctuations in bath water when perturbed (e.g., when the tap or faucet is running), as opposed to the ripples and waves that persist after the perturbation is removed (e.g., when the tap or faucet is turned off). In one case, the water is trying to find its free energy minimum, while the second case solenoidal, divergence free flow is more like the complicated swinging of a frictionless pendulum that neither consumes nor creates energy. On this view, it becomes interesting to characterise the response of the system to perturbation – here, the exogenous inputs provided by the experimental design. Conceptually, we can regard experimental inputs (such as visual afferents to the lateral geniculate) as (active states of) external particles that influence (but are not influenced by) the sensory states of particles at the lowest level. Practically, these experimental inputs were included in the estimation of the coupling parameters that subtend the Jacobian at the lowest scale.

Figure 16 characterises the influence of exogenous, condition-specific effects at different scales in terms of correlations between fluctuations in the eigenstates that can be explained by any of the three experimental inputs (i.e., *visual*, *motion* and *attention*). This analysis suggests that the effect of exogenous inputs can be detected at all scales. For example, at the third scale, eigenmode 17 shows an extremely significant effect of sensory perturbation, dominated by visual *motion*. The associated eigenmode picks out primary visual cortex and extrastriate areas, encroaching upon motion sensitive regions.

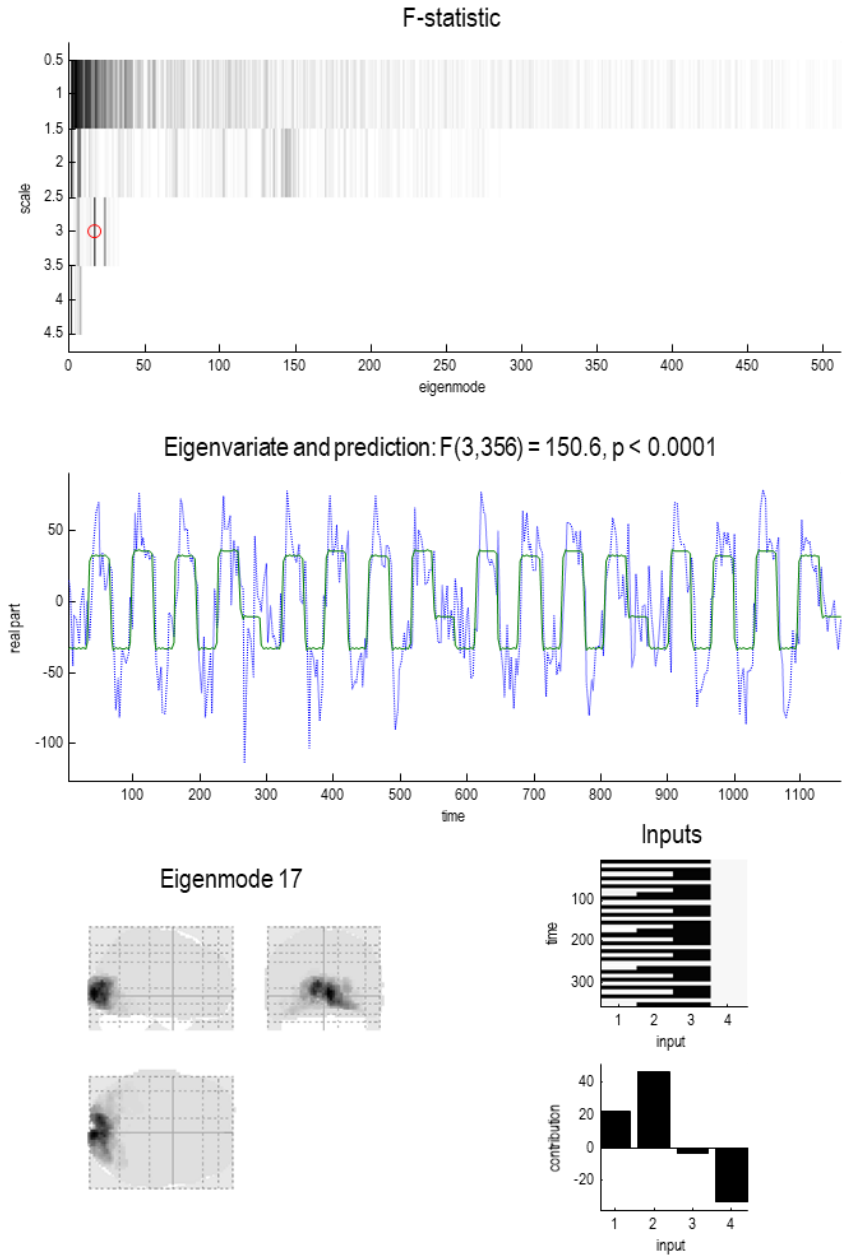

**Figure 16**

**Induced responses**: This figure illustrates the expression of experimental or condition-specific effects at different scales of the particular decomposition. The **top panel** is an unusual form of statistical parametric map; namely, an image of the $F$ statistic, testing for the significance of an effect of any of the three exogenous inputs (i.e., *visual*, *motion* and *attention*). Each row of the $F$ statistic map corresponds to a scale – and comprises the $F$ statistic for each successive particle at that scale. This map shows that the effect of (some linear mixture of) exogenous inputs can be detected at all scales – as evidenced by the dark bars in all four rows. For example, at the third scale, eigenmode 17 shows an extremely significant effect of inputs with an $F$ statistic of over 150 and an exceedingly small $p$-value of less than 0.0001. This eigenmode is shown on the **lower left** in voxel space. Its expression over time – in terms of its real value – is depicted in the middle panel (blue line), with the best fitting prediction based upon exogenous input (green line). This prediction is a contrast (i.e., linear mixture) of the input functions shown in the design matrix on the **lower right**. The coefficients of this contrast are shown below the design matrix; demonstrating that the largest contribution is from the second (*motion*) input. The last column of the design matrix is simply a column of ones. The associated eigenmode picks out primary visual cortex and extrastriate areas, encroaching upon motion sensitive regions in its lateral extremity. The next figure provides a complementary perspective on the effects of inputs, in terms of their first-order kernels or impulse response functions throughout the brain – and over extended periods of time.

Figure 17 provides a complementary and revealing perspective on the effects of sensory perturbation. This figure characterises induced responses in terms of first order Volterra kernels – i.e., impulse response functions – of particles at the first (finest) scale of coarse graining. These kernels are based upon the Jacobian and quantify the effects of changing an input on each eigenmode over time (based on the parameters mediating the influence of experimental inputs on motion of states at the first scale). The maximum intensity projections on the left report the variance attributable to each input, based upon the sum of squared kernels over time (i.e., the auto-covariance function at zero lag under each input). Note that this is a fundamentally different characterisation of brain 'activation' because it is modelling the variance induced by an input that is distributed in space and time through recurrent coupling among brain regions.

In this example, *motion* induces responses in visual and extrastriate; presumably, motion sensitive eigenmodes, while attention has protracted influences on parietal, prefrontal and medial temporal regions; including the frontal eye fields and intraparietal sulcus. Visual input *per se* seems to be expressed preferentially in subcortical systems, including the lateral geniculate but also other subcortical and medial temporal regions. In addition, it appears to selectively engage posterior superior temporal regions in the left hemisphere – often associated with biological motion processing. The interesting aspect of this characterisation is the protracted nature of the kernels – that decay to small values after 100 seconds or so. In effect, this means that although induced responses may be expressed in a regionally specific way almost instantaneously, there are enduring effects that can last for a minute or so, following any exogenous perturbation. Clearly, these effects will be overwritten by ongoing sensory input; however, this suggests that brain systems – and accompanying distributed neuronal responses – have a greater memory than might have been anticipated. Heuristically, this means that I should be able to tell you whether you have 'seen something' in the past minute or so by examining your brain activity at this moment in time.

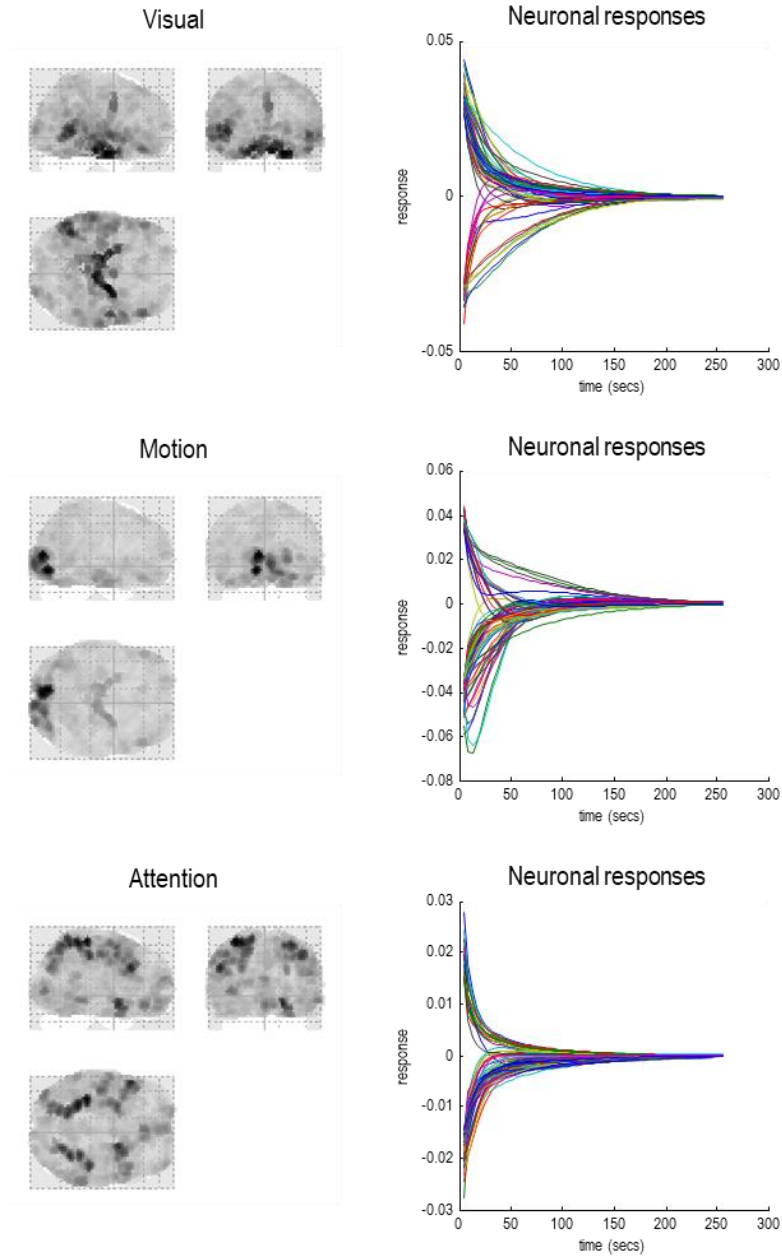

**Figure 17**

**Induced responses over space and time**: this figure characterises induced responses in terms of first order Volterra kernels – i.e., impulse response functions – of particles at the first (finest) scale of coarse graining. Each row corresponds to the three inputs considered (i.e., *visual*, *motion* and *attention* effects). The **left column** shows the expression of these inputs over particles (weighted by the absolute value of their eigenmodes). This effect is the variance attributable to each input (i.e., square of the corresponding kernel, summed over time), shown in the **left row**. These kernels are shown for the 32 particles with the greatest (absolute) magnitude. The key thing to take from these results is that *motion* influences the dynamics of visual and extrastriate; presumably, motion sensitive regions, while attention has protracted influences on parietal, prefrontal and medial temporal regions; including the frontal eye fields and intraparietal sulcus. Interestingly, visual input *per se* seems to be expressed preferentially in subcortical systems, including the lateral geniculate but also other subcortical and medial temporal regions. In addition, visual input appears to selectively engage posterior superior temporal regions in the left hemisphere – often associated with biological motion processing. The more telling aspect of this characterisation is the protracted nature of the kernels – that decay to small values after 100 seconds or so. Notice that these dynamics are supposedly neuronal in nature because we have accommodated haemodynamic convolution at the point of estimating the Jacobian.

## Conclusion

In summary, we have introduced a particular partition that plays the role of a functional parcellation scheme for a system of loosely coupled nonlinear oscillators, such as neuronal populations in the brain. The key aspect of this parcellation scheme is that it can be applied recursively in the spirit of the renormalisation group. This enables one to examine the scale-invariant behaviour of the ensuing spatiotemporal dynamics in a principled way. The numerical analyses above confirm the analytic intuitions that as we move from one scale to the next, there is a progressive slowing and loss of stability of eigenstates associated with each parcel or particle. This manifests as a form of self-organised criticality; in the sense that slow unstable (non-dissipative) eigenmodes supervene on lower scales. Quantitatively speaking, the spatial scale of a particle, its characteristic frequencies and Lyapunov exponents, all fit nicely with empirical observations of (dynamic) functional connectivity within and among large-scale intrinsic brain networks (Liegeois et al., 2019; Northoff, Wainio-Theberge, & Evers, 2019).

Because this paper is a technical (foundational) description of the procedures entailed by the existence of Markov blankets – of Markov blankets – we have focused on the simplest implementation. This means that we started off with linearisation assumptions – and propagated this approximation to higher levels. Clearly, it would be nice to revisit the particular partition using higher-order approximations that retain nonlinearity in the equations of motion; e.g., equation (15). This would require a more careful analysis of the Lyapunov exponents, which would involve integrating the system and averaging the eigenvalues over the ensuing state-dependent Jacobian: the Jacobian becomes a function of states when one includes nonlinearities in the equations of motion. This raises the interesting issue of how to identify the adjacency matrix used to define the Markov blankets. In other words, we need to establish the conditional independences in terms of a zero entry in the Jacobian. However, if the Jacobian is fluctuating over time, over an orbit in state space, then there may be times when the Jacobian element is zero (i.e., zero coupling) and nonzero at other times. Related numerical analyses of nonlinear systems (K Friston, 2013) usually require that the Jacobian is zero over a suitably long period of time, when forming the adjacency matrix in Figure 2. Clearly, this would involve evaluating the Jacobian over all the solutions to the trajectory in state space. This may be a time-consuming but otherwise an interesting exercise.

Technical note

We have already mentioned some limitations and extensions. These include starting off with multivariate characterisations of intrinsic dynamics at the lowest level. As noted above, this is easy to implement by using the first few principal eigenstates, following a singular decomposition of the smallest particles. Another extension is repeating the dynamic causal modelling at each scale, to re-evaluate the Jacobian with suitable high order (i.e., nonlinear) approximations to the equations of motion.

The non-uniqueness of the particular partition is a key practical issue. There is no pretence that there is any unique particular partition. There are a vast number of particular partitions for any given coupled dynamical system. In other words, by simply starting with different internal states – or indeed the number of internal states per particle – we would get a different particular partition. Furthermore, the thresholds used in the elimination of fast dissipative eigenmodes will also change the nature of the partition, leading to more or less inclusive dynamics at the scales above. This latter aspect is probably more defensible in terms of summarising multiscale behaviour; in the sense that we can easily motivate the adiabatic approximation in terms of the relative stability of eigenmodes at a particular level. However, the number of internal states to consider – and how to pick them – introduces a more severe form of non-uniqueness. In this paper, we used the state that was maximally coupled to other states as the internal state of the next particle. This was based upon the graph Laplacian of the adjacency matrix at the appropriate scale. This is a sensible but somewhat arbitrary definition of an internal state and speaks to the point that there are a multitude of particular partitions – and implicit Markov blankets – that could be used. There are two ways that one could handle this non-uniqueness. One would be to embrace it and focus on the statistics of characterisations over different particular partitions and look for scaling behaviours that are conserved over partitions. The alternative is to think about a unique particular partition and how this would be identified. This as an outstanding issue; namely, what is the 'best' particular partition and, indeed, is the notion of the best partition appropriate?

Another important caveat is the fact that we have predicated the illustrative analyses in this paper on a single-subject dataset acquired under an experimental activation paradigm. We chose this dataset because it has been used to illustrate previous developments of dynamic causal modelling. Conceptually, this means that the particular partition is specific to this subject and the subject's responses to the attentional paradigm (summarised in Figure 17). Because this paradigm introduced context sensitive or condition specific changes in effective connectivity, it was designed to change the Jacobian over different periods of stimulation (e.g., attentional modulation of coupling between visual motion areas and early visual cortex). We did not attempt to model these effects here – this would require the nonlinear modelling mentioned above. If this modelling was to second order, we would end up with a bilinear form for equation (2), which is the basis of most DCM analyses of fMRI data. This speaks to the fact that the parcellation scheme may not produce the same results when applied to a different paradigm. In turn, means that there is further work to be done in terms of finding a particular partition that accommodates variability in functional anatomy. In theory, this would probably be best addressed using a generative model. In other words, assuming one underlying sparse Jacobian at any given scale and then add random effects, so that it could be used to explain multiple paradigms or subjects. Having said this, the current analyses can be taken as proof of principle that this sort of multiscale decomposition can be applied to empirical neuroimaging timeseries – and leads to the same phenomenology reported in a functional connectivity literature.

## Appendix: Markov blankets for random dynamical systems

Formally, the definition of Markov blankets, in terms of dynamical (i.e., causal) influences, is a little more delicate than their definition given a probabilistic graphical model (i.e., conditional dependencies). This is because the conditional dependencies among the states of a dynamical system are those that obtain at nonequilibrium steady state, which depends upon dynamical coupling among states in a nontrivial way. The aim here is to identify sufficient conditions that render subsets of states conditionally independent of each another – so that they can be distinguished in a statistical sense.

**Definition** (*dissipative partition*): a dissipative partition is a partition into external, blanket (i.e., sensory and active) and internal states, where internal and external states do not influence each other – and one or more subset of states is dissipative, i.e., the leading diagonal elements of the associated Jacobian are large and negative.

**Lemma** (*Markov blankets)*: The sensory and active states of a *dissipative* partition constitute a Markov blanket $b = (s, a)$ that renders external and internal states conditionally independent:

$$(\mu \perp \eta)|b \Leftrightarrow p(\mu, \eta|b) = p(\mu|b)p(\eta|b) \tag{22}$$

**Proof**: at nonequilibrium steady state, the following solution to the Fokker Planck equation holds (Ao, 2004; Qian & Beard, 2005):

$$f(x) = (Q - \Gamma)\nabla \Im(x) \tag{23}$$

Here, $\mathfrak{I}(x) = -\ln p(x)$ is surprisal or self-information and the antisymmetric (skew) matrix $Q = -Q^\dagger$ mediates solenoidal flow. The positive definite matrix $\Gamma \propto I$ is a diffusion tensor describing the amplitude of random fluctuations. In this (Helmholtz) decomposition, the flow $f(x)$ can be decomposed into dissipative gradient flows $-\Gamma\nabla\mathfrak{I}$ and divergence free or solenoidal flow $Q\nabla\mathfrak{I}$. Differentiating, with respect to the states, evinces the relationship between the flow – specified by a Jacobian $J = \nabla f(x)$ – and conditional independencies – specified by a Hessian $H = \nabla^2\mathfrak{I}$:

$$\begin{aligned}
\nabla f(x) &= (Q - \Gamma)\nabla^2\mathfrak{I}(x) \Rightarrow \\
J(x) &= (Q - \Gamma)H(x) \Rightarrow \\
H(x) &= -(\Gamma - Q)^- J(x) \approx -(\Gamma + Q)J(x)
\end{aligned} \tag{24}$$

Here, the coupling is encoded by the Jacobian[9]. For example, if the Jacobian encoding the coupling between external and internal states is zero, we can express the flow of internal states as a function of, and only of, particular states:

$$J_{\mu\eta} = \nabla_\eta f_\mu(x) = 0 \Rightarrow f_\mu(x) = f_\mu(\pi) \tag{25}$$

Similarly, the Hessian or curvature matrix encodes conditional dependencies, in the sense that if the corresponding submatrix is zero, internal and external states are conditionally independent:

$$H_{\mu\eta} = \nabla_{\mu\eta}\mathfrak{I}(x) = 0 \Rightarrow \mathfrak{I}(\mu|b,\eta) = \mathfrak{I}(\mu|b) \Rightarrow (\mu \perp \eta)|b \tag{26}$$

Equation (24) shows that the amplitude of random fluctuations and solenoidal coupling play a key role in relating dynamic coupling and conditional dependencies. The solenoidal components are especially important in the setting of nonequilibrium steady state. Indeed, on one reading of nonequilibrium dynamics, the very presence of solenoidal flow is sufficient to break detailed balance – and preclude any equilibria in the conventional (statistical mechanics) sense (Ao, 2005; Kwon & Ao, 2011; Seifert, 2012; Zhang et al., 2012).

---

[9] The approximate equality follows from a first-order Taylor expansion of the inverse of a mixture of matrices.

The symmetry of the Hessian matrix places linear constraints on the solenoidal coupling (Qian & Beard, 2005); where, dropping the dependency on $x$ for simplicity:

$$
\begin{aligned}
(Q-\Gamma)^{-1} J &= \Pi = \Pi^T = J^T (Q-\Gamma)^{-T} \\
&\Rightarrow \\
JQ + QJ^T &= \Gamma J^T - J\Gamma \Rightarrow \\
vec(Q) &= (I \otimes J + J \otimes I)^- vec(\Gamma J^T - J\Gamma)
\end{aligned}
\tag{27}
$$

These constraints mean that the solenoidal flow can be expressed as a function of the Jacobian and the amplitude of random fluctuations, as shown in the last equality of equation (27). In turn, this means we can express the Hessian, encoding conditional independencies, as a function of the Jacobian. For example, in a system with one external, blanket and active state, substituting equation (27) into equation (24) gives[10]:

$$
H(x) = \begin{bmatrix}
\dfrac{64\kappa^7 + \cdots}{64\kappa^6 + \cdots} & \dfrac{-32\kappa^6(J_{b\eta} + J_{\eta b}) + \cdots}{64\kappa^6 + \cdots} & \dfrac{16\kappa^5(J_{b\eta}J_{b\mu} - J_{\eta b}J_{\mu b}) + \cdots}{64\kappa^6 + \cdots} \\[2ex]
\dfrac{-32\kappa^6(J_{b\eta} + J_{\eta b}) + \cdots}{64\kappa^6 + \cdots} & \dfrac{64\kappa^7 + \cdots}{64\kappa^6 + \cdots} & \dfrac{-32\kappa^6(J_{b\mu} + J_{\mu b}) + \cdots}{64\kappa^6 + \cdots} \\[2ex]
\dfrac{16\kappa^5(J_{b\eta}J_{b\mu} - J_{\eta b}J_{\mu b}) + \cdots}{64\kappa^6 + \cdots} & \dfrac{-32\kappa^6(J_{b\mu} + J_{\mu b}) + \cdots}{64\kappa^6 + \cdots} & \dfrac{64\kappa^7 + \cdots}{64\kappa^6 + \cdots}
\end{bmatrix}
$$

$$
J(x) = \begin{bmatrix}
J_{\eta\eta} - \kappa & J_{\eta b} & \\
J_{b\eta} & J_{bb} - \kappa & J_{b\mu} \\
& J_{\mu b} & J_{\mu\mu} - \kappa
\end{bmatrix}, \quad
\Gamma = \begin{bmatrix}
I & & \\
& I & \\
& & I
\end{bmatrix}
\tag{28}
$$

Here, the elements of the Hessian have been expressed as rational functions (ratios of polynomials) of $\kappa > 0$, retaining the leading orders. These functions have horizontal and linear asymptotes, such that in the limit of dissipative flows, we have:

---

[10] Note that we have eliminated the amplitude of random fluctuations by assuming, without loss of generality, the states have been suitably scaled to render $\Gamma = I$. Furthermore, to simplify the (symbolic) maths, we have used the Taylor approximation in equation (24).

$$\lim_{\kappa \to \infty} H(x) = \begin{bmatrix} \kappa & -\frac{1}{2}(J_{b\eta} + J_{\eta b}) & 0 \\ -\frac{1}{2}(J_{b\eta} + J_{\eta b}) & \kappa & -\frac{1}{2}(J_{b\mu} + J_{\mu b}) \\ 0 & -\frac{1}{2}(J_{b\mu} + J_{\mu b}) & \kappa \end{bmatrix} \Rightarrow$$

(29)

$$\lim_{\kappa \to \infty} H_{\eta\mu}(x) = 0 : \forall x \Rightarrow (\mu \perp \eta) \,|\, b$$

In short, for sufficiently dissipative systems, the linear constraints on solenoidal flow ensure conditional independence between internal and external states, given blanket states. □

The above proof assumed single states; however, the results can be generalised using numerical analyses (or symbolic maths) for high dimensional systems. An example is presented in Figure 1

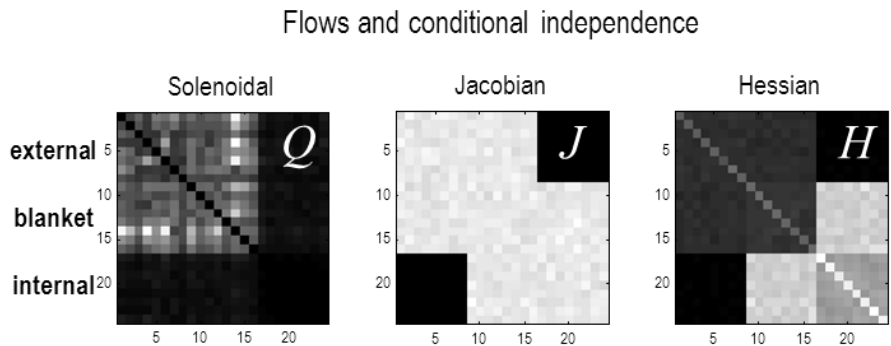

**Figure 18**

**Dissipation and conditional independence**: numerical analyses that show the conditional independence between internal and external states depends upon dissipation, as quantified by the average value of the leading diagonal Jacobians. In this example, a system with 24 states was divided equally into external, blanket, and internal states. The panels above report the variance of the solenoidal term, the Jacobian and Hessian, based on 512 random samples where each element of the Jacobian was sampled from a unit Gaussian distribution and values of 4, 4 and 32 were added to the leading diagonal for the external, blanket and internal states, respectively. The black patches on the lower left (and upper right) shows that an absence of coupling in the Jacobian – between the external and internal states – precludes solenoidal coupling and renders the external and internal states conditionally independent.

# Software note

The software producing the figures in this figure are available as part of the academic software SPM. They can be accessed by invoking DEM graphical user interface and selecting the **DCM and blankets** button (DEMO_DCM_MB.m). Please see https://www.fil.ion.ucl.ac.uk/spm/.

**Acknowledgements**

K.J.F. was funded by a Wellcome Trust Principal Research Fellowship (Ref: 088130/Z/09/Z). E.D.F. was supported by a King's College London Prize Fellowship. T.S.Z. was supported by the Cognitive Sciences and Technologies Council of Iran for an international research visit. AR was funded by Australian Research Council (Refs: DE170100128 and DP200100757) and National Health Medical Research Council (Ref:APP1194910).

**Declaration**

The authors have no conflicts of interest or declarations